\documentclass[aps,prl,reprint,twocolumn,showpacs,superscriptaddress]{revtex4-1}

\usepackage[utf8]{inputenc}
\usepackage[english]{babel}  % english language

\usepackage{times}
\usepackage{amssymb,amsmath,amsfonts,amsthm}
\usepackage{mathtools}
\usepackage{verbatim}
\usepackage{enumerate}
\usepackage{graphicx}
\graphicspath{{pics/}}
\usepackage[colorlinks=true,linkcolor=blue,citecolor=blue,urlcolor=blue]{hyperref}
\usepackage{bbm}
\usepackage{booktabs}

\usepackage{subfigure}
\usepackage{color}
\definecolor{dred}{rgb}{.8,0.2,.2}
\definecolor{ddred}{rgb}{.8,0.5,.5}
\definecolor{dblue}{rgb}{.2,0.2,.8}
\definecolor{dgreen}{rgb}{.2,0.5,.2}

\theoremstyle{plain}
\theoremstyle{definition}

\newcommand{\bra}[1]{\mbox{$\langle #1|$}}
\newcommand{\ket}[1]{\ensuremath{|#1\rangle}}

\newcommand{\be}{\begin{equation}}
	\newcommand{\ee}{\end{equation}}

\usepackage{tikz}
\usepackage{braids}
\usepackage{xifthen}

\usepackage{euscript}

\begin{document}

\title{Entanglement-Enhanced Quantum Metrology in Colored Noise
by Quantum Zeno Effect}

\author{Xinyue Long}
	\thanks{These authors contributed equally to this work.}
\affiliation{Shenzhen Institute for Quantum Science and Engineering and Department of Physics, Southern University of Science and Technology, Shenzhen 518055, China}

\author{Wan-Ting He}
	\thanks{These authors contributed equally to this work.}
\affiliation{Department of Physics, Applied Optics Beijing Area Major Laboratory, Beijing Normal University, Beijing 100875, China}

\author{Na-Na Zhang}
	\thanks{These authors contributed equally to this work.}
	\affiliation{School of Optoelectronic Engineering, Chongqing University of Posts and Telecommunications, Chongqing 400065, China}
\affiliation{Department of Physics, Applied Optics Beijing Area Major Laboratory, Beijing Normal University, Beijing 100875, China}

%\author{Ru-Qiong Deng}
%\affiliation{Department of Physics, Applied Optics Beijing Area Major Laboratory, Beijing Normal University, Beijing 100875, China}

\author{Kai Tang}
\affiliation{Shenzhen Institute for Quantum Science and Engineering and Department of Physics, Southern University of Science and Technology, Shenzhen 518055, China}

\author{Zidong Lin}
\affiliation{Shenzhen Institute for Quantum Science and Engineering and Department of Physics, Southern University of Science and Technology, Shenzhen 518055, China}

\author{Hongfeng Liu}
\affiliation{Shenzhen Institute for Quantum Science and Engineering and Department of Physics, Southern University of Science and Technology, Shenzhen 518055, China}

\author{Xinfang Nie}
\affiliation{Shenzhen Institute for Quantum Science and Engineering and Department of Physics, Southern University of Science and Technology, Shenzhen 518055, China}
 \affiliation{Guangdong Provincial Key Laboratory of Quantum Science and Engineering, Southern University of Science and Technology, Shenzhen 518055, China}

 \author{Guanru Feng}
\affiliation{Shenzhen SpinQ Technology Co., Ltd., Shenzhen, China}

\author{Jun Li}
\affiliation{Shenzhen Institute for Quantum Science and Engineering and Department of Physics, Southern University of Science and Technology, Shenzhen 518055, China}
 \affiliation{Guangdong Provincial Key Laboratory of Quantum Science and Engineering, Southern University of Science and Technology, Shenzhen 518055, China}

\author{Tao Xin}

\affiliation{Shenzhen Institute for Quantum Science and Engineering and Department of Physics, Southern University of Science and Technology, Shenzhen 518055, China}
 \affiliation{Guangdong Provincial Key Laboratory of Quantum Science and Engineering, Southern University of Science and Technology, Shenzhen 518055, China}

\author{Qing Ai}
\email{aiqing@bnu.edu.cn}
\affiliation{Department of Physics, Applied Optics Beijing Area Major Laboratory, Beijing Normal University, Beijing 100875, China}

\author{Dawei Lu}
\email{ludw@sustech.edu.cn}
\affiliation{Shenzhen Institute for Quantum Science and Engineering and Department of Physics, Southern University of Science and Technology, Shenzhen 518055, China}
 \affiliation{Guangdong Provincial Key Laboratory of Quantum Science and Engineering, Southern University of Science and Technology, Shenzhen 518055, China}

%\author{Xinyue Long,$^{1,*}$
%Wan-Ting He,$^{2,*}$
%Na-Na Zhang,$^{3,2,}$\footnote{These authors contributed equally to this work.}
%Ru-Qiong Deng,$^{2}$
%Kai Tang,$^{1}$
%Zidong Lin,$^{1}$
%Hongfeng Liu,$^{1}$
%Xinfang Nie,$^{1}$
%Guanru Feng,$^{4}$
%Jun Li,$^{1}$
%Tao Xin,$^{1}$
%Qing Ai,$^{2,}$\footnote{aiqing@bnu.edu.cn}
%and Dawei Lu,$^{1,4}$\footnote{ludw@sustech.edu.cn}}

%\address{Shenzhen Institute for Quantum Science and Engineering and Department of Physics, Southern University of Science and Technology, Shenzhen 518055, China \\
% $^{2}$Department of Physics, Applied Optics Beijing Area Major Laboratory, Beijing Normal University, Beijing 100875, China \\
%$^{3}$ School of Optoelectronic Engineering, Chongqing University of Posts and Telecommunications, Chongqing 400065, China \\
%$^{4}$ Shenzhen SpinQ Technology Co., Ltd., Shenzhen, China\\
%$^{5}$ Guangdong Provincial Key Laboratory of Quantum Science and Engineering, Southern University of Science and Technology, Shenzhen 518055, China\\
%}

\date{\today}

\begin{abstract}
  In open quantum systems, the precision of metrology inevitably suffers from the noise. {In Markovian open quantum dynamics, the precision can not be improved by using entangled probes although the measurement time is effectively shortened.} However, it was predicted over one decade ago that in a non-Markovian one, the error can be significantly reduced by the quantum Zeno effect (QZE) [Chin, Huelga, and Plenio, Phys. Rev. Lett. \textbf{109}, 233601 (2012)]. In this work, we apply a recently-developed quantum simulation approach to experimentally verify that entangled probes can improve the precision of metrology by the QZE. Up to $n=7$ qubits, we demonstrate that the precision has been improved by a factor of $n^{1/4}$, which is consistent with the theoretical prediction. Our quantum simulation approach may provide an intriguing platform for experimental verification of various quantum metrology schemes.
\end{abstract}
\maketitle

%\section*{Introduction}

%1.There are many interesting phenomena in non-Markovian environment.
%
%2. Quantum metrology in different environments: Markovian vs non-Markovian
%environment. But it is difficult to experimentally verify the theoretical
%proposal.
%
%3. Our quantum simulation approach is theoretically exact, fully controllable,
%and practically efficient.
%
%4. We apply our quantum simulation approach to experimentally verify
%the quantum metrology scheme in the non-Markovian environment.

\textit{Introduction.}---Quantum metrology utilizes quantum entanglement and coherence to enhance the measurement precision over classical metrology \cite{Caves1981,Giovannetti2004a,Pezze2018}. In the absence of noise, the precision of classical metrology scales as $n^{-1/2}$, which is limited by the central-limit theorem with $n$ being the number of resources implemented in the measurements. In contrast, quantum metrology can reach the Heisenberg limit, scaling as $n^{-1}$. However, in practice, any quantum system is inevitably subject to decoherence which results from the interaction with the environment \cite{Breuer2002}. Due to decoherence, there will be no improvement in the measurement precision over classical metrology, although the measurement time may be effectively shortened \cite{Huelga1997}. {In order to overcome this shortcoming, various methods have been put forward, e.g., squeezing \cite{Ma2011b}, purification \cite{Yamamoto2022}, and one-way quantum-computing-based teleportation \cite{Matsuzaki2018}.} It was shown that instead of maximally entangled states, partially entangled initial states may reduce the error to some extent \cite{Huelga1997}. When the system interacts with a specific bath with a band structure, ideal precision may be retrieved as a result of the existence of the bound state \cite{Bai2019a,Ai2010}. {When there is a correlation between the baths for individual qubits, an auxiliary qubit can be introduced to reach the Heisenberg limit \cite{Monz2011,He2021,Kukita2021}.} Interestingly, the nonlinear interaction between the system and the physical quantity to be measured is used to surpass the Heisenberg limit \cite{Boixo2007,Nie2015,Choi2008,Roy2008,Napolitano2011}. However, the nonlinear interaction may not be easily realized in practical measurements.

Then, a question naturally comes to our minds: Can we figure out some practical method to improve the measurement precision in the presence of noise? It is shown that there are generally three stages in open quantum dynamics, including the well-known exponential decay in the intermediate stage \cite{Nakazato1996}. {However, in the first stage, the open quantum system decays in a Gaussian way, where the quantum Zeno effect (QZE) \cite{Ai2010a,Ai2013,Harrington2017,Virzi2022,Kiilerich2015,Do2019,Burgarth2014} happens.} {It has been pointed out that due to the QZE, metrology using maximally entangled states is superior to the one using product states by a factor of $n^{1/4}$ \cite{Chin2012,Matsuzaki2011}.} On the other hand, recently we have theoretically proposed and experimentally demonstrated in a nuclear magnetic resonance (NMR) platform, to simulate the quantum
dynamics for various Hamiltonians and spectral densities \cite{Wang2018b,Zhang2021,Chen2022}. In these works, the bath-engineering technique \cite{Soare2014a,Soare2014,Khaneja2005b,Li2017} enables a theoretically exact, fully controllable, and practically
efficient quantum simulation \cite{Buluta2009,Georgescu2014} approach. {Using the approach, we show that the non-Markovianity of the open quantum dynamics should be essentially characterized from both aspects of global and local points of view, e.g., quantum mutual information v.s. quantum Fisher information flows \cite{Chen2022,Luo2012,Lu2010,Altherr2021}.} It is this theoretically exact and fully controllable characteristic that enables us to experimentally verify the quantum metrology scheme proposed in Refs.~\cite{Chin2012,Matsuzaki2011}, which requires homogeneity of the qubits.

\begin{figure}
\includegraphics[scale=0.75]{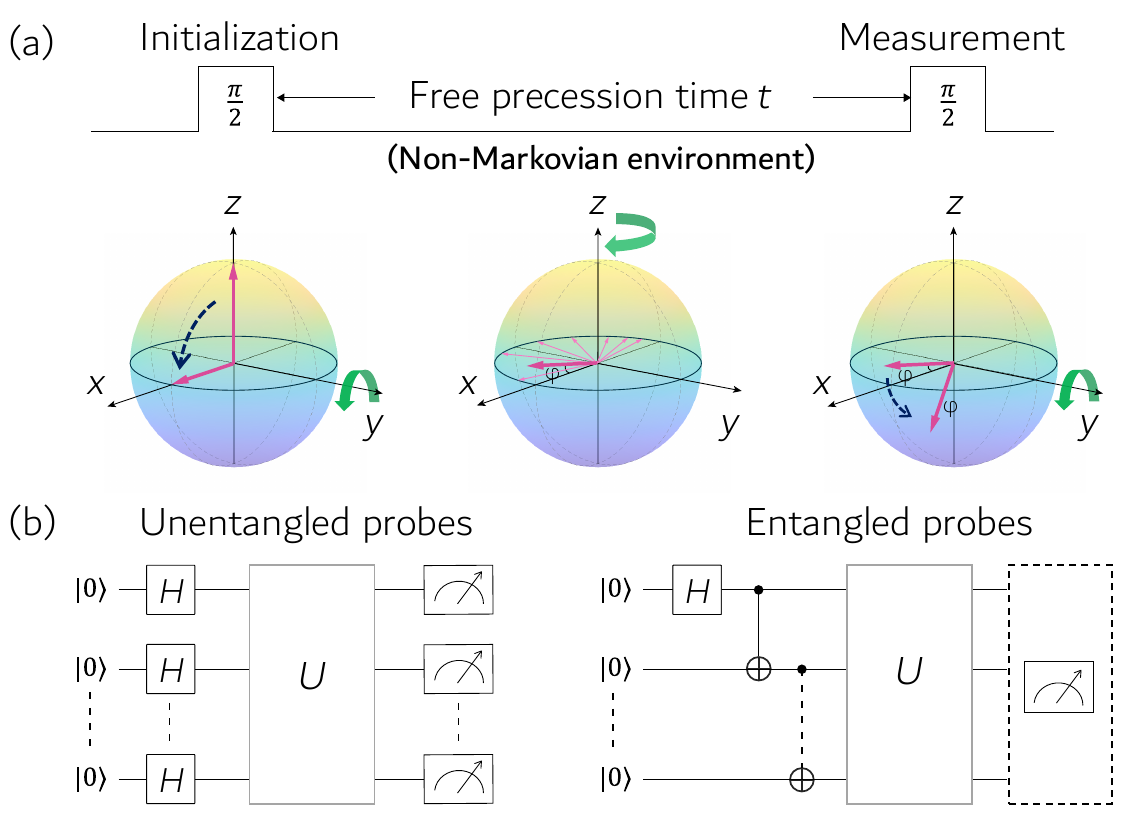}\caption{(a) Single-qubit Ramsey experiment for quantum metrology under non-Markovian noise. The free precession is governed by non-Markovian dynamics, which can be simulated using the ensemble-averaging technique. This averaging can be visualized in the Bloch sphere: A large number of identical initial probes undergo phase diffusion owing to their different modulated Hamiltonians, leading to an attenuation in amplitude for the overall spin signal. This attenuation gives rise to the error in quantum-metrology experiments. (b) Quantum circuits for multi-qubit Ramsey experiment under non-Markovian noise with unentangled and entangled probes, respectively. The evolution $U$ is the simulation of non-Markovian dynamics. For the case of entangled probes, the readout is typically some collective measurement \cite{Ma2011b,Huelga1997}.}
\label{fig:scheme}
\end{figure}

\textit{Protocol.}---{The bath-engineering technique offers a way to engineer arbitrary environments by modulating the control field \cite{Soare2014a,Soare2014}. By applying a time-dependent magnetic field to the total system, we can artificially add a spectral characteristic noise, so as to simulate an arbitrary noisy environment.} The Hamiltonian of the total system can be written as
$H(t)=H_\textrm{S}+H_\textrm{SB}(t)$,
{where $H_\textrm{S}=\sum_{m}\varepsilon_m\vert m\rangle\langle m\vert+\sum_{m\neq n}J_{mn}\vert m\rangle\langle n\vert$ is the
system Hamiltonian, and $H_\textrm{SB}(t)$ is the noise Hamiltonian to simulate the system-bath couplings.} When simulating the pure-dephasing noise, cf. Fig.~\ref{fig:scheme}(a), the noise Hamiltonian is $H_\textrm{SB}(t)=\sum_m\beta_m(t)\vert m\rangle\langle m\vert$.
$\beta_{m}(t)$'s are stochastic errors generated by performing amplitude
and phase modulations on a carrier \cite{Wang2018b,Zhang2021}
\begin{equation}
\beta_{m}(t)=\alpha_{z}\sum_{j=1}^{J}\omega_{j}F(\omega_{j})\cos\left(\omega_{j}t+\psi_{j}^{(m)}\right),
\label{eq:beta}
\end{equation}
where $\alpha_{z}$ is the amplitude of the dephasing noise, $\omega_\textrm{b}$ ($J\omega_\textrm{b}$) is the base (cutoff) frequency with $\omega_{j}=j\omega_\textrm{b}$, $F(\omega_{j})$ is the modulation function of a specific spectral characteristic noise, and
$\psi_{j}^{(m)}$'s are a set of random numbers. For a noise of Gaussian type, the average over the ensemble is equivalent to the average over the time \cite{Goodman2015}. {
The ensemble-averaged decoherence factor reads
$\Gamma(t)  =\alpha_{z}^{2}\sum_{j=1}^{J}F(\omega_{j})^{2}\sin^{2}\frac{\omega_{j}t}{2}$,
as determined by the power spectral density of
the noise, which is the Fourier transform of the second-order
correlation function $\langle \beta_{m}(t+\tau)\beta_{m}(t)\rangle$ \cite{Soare2014a,Soare2014}.
}

We consider a system Hamiltonian $H_\textrm{S}=\sum_{k=1}^{n}\omega_{0}\sigma_{k}^{z}/2$,
which is subject to a pure-dephasing noise \cite{Chin2012}. We present an illustration for the case with a single qubit in Fig.~\ref{fig:scheme}(a). Initially, the qubit is prepared at $(\vert0\rangle+\vert1\rangle)/\sqrt{2}$. Afterwards, it unitarily evolves at the equator of the Bloch sphere with rate $\omega_{0}+\beta_1(t)$. In order to mimic the pure dephasing, we generate a large number of realizations with the same initial state and obtain the final result by averaging over the random ensemble with different rate $\beta_1(t)$ after a second $\pi/2$ pulse.
We assume a Drude-Lorentz spectral density, i.e., $J_{k}(\omega)=J(\omega)=2\lambda\gamma\omega/(\omega^{2}+\gamma^{2})$,
where $\lambda$ is the reorganization energy and $\gamma$ is the
relaxation rate. A generic spectral density
can be decomposed into a summation of Lorentzian forms. The modulation function then becomes $F(\omega_{j})=[\lambda\gamma\omega_{0}\coth(\beta\omega_{j})/(\omega_{j}^{3}+\gamma^{2}\omega_{j})]^{1/2}$.
{The quantum dynamics of the system is exactly described by the time-local master equation \cite{Breuer2002,Tao2020}
\begin{equation}
\frac{\partial}{\partial t}\rho(t)=-i[H_\textrm{S},\rho]+\sum_{k}\gamma_{k}(t)(\sigma^{z}_{k}\rho\sigma^{z}_{k}-\rho).
\label{eq:ME}
\end{equation}
Note that $\gamma_{k}(t)$ is explicitly dependent on time. According
to the bath-engineering technique, the ensemble-averaged decoherence
rate is
$\gamma_{k}(t)=\frac{\alpha_z^{2}}{2}\sum_{j=1}^{J}\omega_{j}F(\omega_{j})^{2}\sin(\omega_{j}t)$.
}

%\section*{Quantum Metrology}
In Fig.~\ref{fig:scheme}(b), we present the quantum circuit for the quantum metrology scheme for the product and entangled states, respectively. For the former case, the probes are prepared at $[(\vert0\rangle+\vert1\rangle)/\sqrt{2}]^{\otimes n}$ and then evolve under the unitary evolution $U=\exp(-i\int_0^t d\tau[H_\textrm{S}+\sum_k\beta_k(\tau)\sigma_k^z])$, which will not induce entanglement between different qubits but mimic the noise. Then, we perform the individual measurements on each qubit, respectively, since all qubits are disentangled. However, for the latter case, since the probes are initialized at the maximally entangled state $(\vert0\rangle^{\otimes n}+\vert1\rangle^{\otimes n})/\sqrt{2}$, we only perform a collective measurement after the same $U$. {
For open quantum dynamics governed by the master equation~(\ref{eq:ME}), the variances of the measured frequency for the unentangled and entangled probes are respectively
 $\delta\omega^2_0\vert_u=\exp[2\Gamma(t_u)]/(nTt_u)$ and $\delta\omega^2_0\vert_e=\exp[2n\Gamma(t_e)]/(n^2Tt_e)$,
where $\Gamma(t)=\int_0^t\gamma_k(t^\prime)dt^\prime$ is the decoherence factor for a single qubit, and the subscript $u$ ($e$) refers to the unentangled (entangled) state. The optimal times to perform the measurement are determined by
$2t_u\gamma_k(t_u)=1$ and
$2nt_e\gamma_k(t_e)=1$, respectively.
When the decoherence rate $\gamma_{k}(t)=c$ is time-independent, the variances $\delta\omega^2_0=2ce/(nT)$ for both initial states are the same, although the optimal times are different, i.e., $t_u=1/(2c)$ v.s. $t_e=1/(2nc)$. However, when the QZE occurs, i.e., $\gamma_{k}(t)=2ct$, the variance for the unentangled probes is inferior to that for the entangled ones, as $\delta\omega^2_0\vert_u=2\sqrt{ce}/(nT)>\delta\omega^2_0\vert_e=2\sqrt{nce}/(n^2T)$
with $t_u=(4c)^{-1/2}>t_e=(4nc)^{-1/2}$.} Hereafter, we shall give an introduction to the experimental details.

\begin{figure*}
\includegraphics[scale=1]{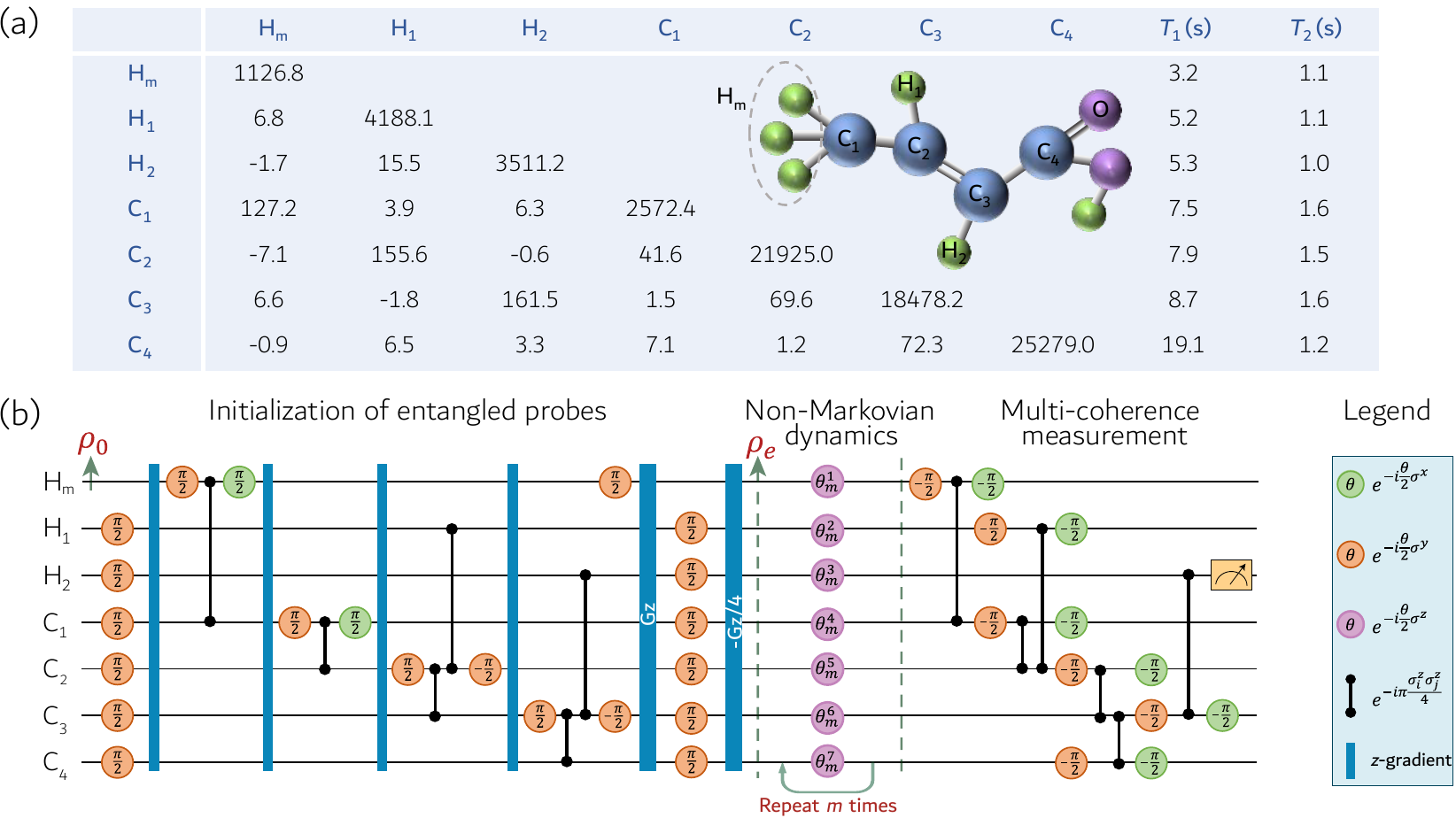}\caption{(a) Molecular structure and relevant parameters of the 7-qubit NMR processor. Chemical shifts (diagonal, Hz), scalar coupling strengths (off-diagonal, Hz), and relaxation times ($T_1$ and $T_2$) are all listed in the table. (b) Pulse sequence for the 7-qubit Ramsey experiment using the entangled probe under non-Markovian noise. To simulate the non-Markovian dynamics, the rotating angles $\theta_m^j$ are carefully designed by ensemble-averaging modulation. A step-by-step description of the sequence can be found in SM \cite{sup}.}
\label{fig:circuit}
\end{figure*}

%By the above quantum simulation approach, we simulate the Ramsey experiment for $n=7$ qubits, as shown in Fig~\ref{fig:circuit}. Compared to the unentangled case, the population dynamics manifests an enhanced decay due to the entangled state. By varying the number of qubits, we can compare the ratio $r=\delta\omega_0\vert_u/\delta\omega_0\vert_e$ to estimate the improvement of measured frequency by using the entangled probes. The variances of the entangled probes are generally lower than those of the unentangled ones. And the experimentally-observed $r$ approaches the theoretical prediction, which scales as $r=n^{1/4}$.

%1. Show the result of experiment.
%
%2. Discuss the result and analyze the error.

%\begin{widetext}

%\begin{figure}
%\includegraphics[width=18cm]{Q2}\caption{Experimental results for 2 qubits.}
%\end{figure}
%
%\begin{figure}
%\includegraphics[width=18cm]{Q3}\caption{Experimental results for 3 qubits.}
%\end{figure}
%
%\begin{figure}
%
%\includegraphics[width=18cm]{fig2}\caption{Experimental results for 4 qubits.}
%
%\end{figure}

%\end{widetext}
%\begin{figure}
%\includegraphics[width=9cm]{fig3}\caption{Relative ration vs the number of qubits.}
%\end{figure}

\emph{Experiment.}---The experiments are performed on a Bruker 600 MHz spectrometer at room temperature. {The sample is $^{13}$C-labeled trans-crotonic acid dissolved in $d_6$-acetone, which consists of three $^1$H and four $^{13}$C nuclear spins, forming a 7-qubit quantum processor.} The molecular structure of the sample is shown in Fig.~\ref{fig:circuit}(a). The internal Hamiltonian of this system can be described as
\begin{equation}
	\mathcal{H}_{\rm{NMR}}=-\sum_{i=1}^7 \frac{\omega_i}{2} \sigma^z_{i}+\sum_{i<j,=1}^7\frac{\pi}{2} J_{ij}\sigma^z_{i}\sigma^z_{j},
\end{equation}
where $\omega_i/2\pi$ is the Larmor frequency of the $i$-th spin, and $J_{ij}$ is the scalar coupling strength between spins $i$ and $j$. The corresponding parameters are listed in Fig.~\ref{fig:circuit}(a), as well as the relaxation times $T_1$ and $T_2$.

Overall, the experiment aims to observe the enhancement of quantum metrology with entangled probes under the non-Markovian circumstance. This is demonstrated using the Ramsey magnetometry, which typically contains three major stages: (i) preparation of initial probe states, (ii) accumulation of an energy-splitting dependent phase, and (iii) measurement of that phase. In the following, we will describe each stage in detail.

(i) Prepare the two types of probe states: the unentangled probe $\ket{\psi_0}_u=[(\vert0\rangle+\vert1\rangle)/\sqrt{2}]^{\otimes n}$, and the maximally entangled probe $\ket{\psi_0}_e = (\ket{0}^{\otimes n}+\ket{1}^{\otimes n})/\sqrt{2}$.
For the unentangled probe, the processor is initialized to the pseudo-pure state $\ket{0}^{\otimes n}$ (up to an identity matrix) from the thermal equilibrium state $\rho_{0}$ using the cat-state method  \cite{Knill2000a}. Followed by a 2-ms-shaped pulse to perform  a collective single-qubit rotation $R_y(\pi/2)$ on all qubits, the system is prepared into $\ket{\psi_0}_u$ with over $99\%$ fidelity. The pulse sequence can be found in Supplemental Material (SM) \cite{sup}.

{For the entangled probe, we directly start from the thermal equilibrium $\rho_{0}$, and eventually prepare the state $\rho_e=\ket{0}^{\otimes n}\bra{1}+\ket{1}^{\otimes n}\bra{0}$ with the aid of nearest-neighbour scalar couplings and gradient-echo techniques \cite{sup}, as shown in Fig.~\ref{fig:circuit}(b). Despite not the exact form of $\ket{\psi_0}_e$, the dynamics of $\rho_e$ for quantum metrology purpose is the same since the frequency information is just encoded in the phase accumulation of coherent terms \cite{sup}.} In addition, the creation of $\rho_e$ avoids regular initialization of the pseudo-pure state, which reduces the sequence complexity remarkably.

%(ii) {The Hamiltonian of the energy splitting $\omega_0$ can be written as $H_s= \frac{\omega_0 }{2}\sum_{i=1}^n\sigma^z_{i}$. Let the probes precess unperturbedly for a duration $t$ under the non-Markovian noise, and the phase accumulates of energy splitting is $\phi=\omega_0t$. To simulate the non-Markovian circumstance, we employ the aforementioned bath-engineering technique to engineer arbitrary environments by modulating the control field. Explicitly, we use the time-dependent Hamiltonian in each experiment which has the form $H_{SB}^m(t)=\sum_{i=1}^n\beta^m_{i}(t)\sigma^z_{i}$ to modulate approaching the non-Markovian noise.
%Here $\beta_{i}^m(t)$'s are stochastic errors subject to Drude-Lorentz spectral density as shown in Eq.~(\ref{eq:beta}) and $m$ is the sample index to modulate the non-Markovian via ensemble average noise for the $m$-th experiment. Therefor, the total Hamiltonian is $H_s+H_{SB}(t)$. In the experiment, the evolution of this Hamiltonian corresponds to collective single-qubit rotations about the $z$-axis, which can be easily implemented in most physics systems. In addition, this total Hamiltonian can be discretized into time-independent Hamiltonians with $k=10^3$ slices \cite{sup}. Regarding the experiment, we set $\omega_0=10$~kHz, and average the evolution under the Hamiltonian $H_s+H_{SB}(t)$ by $m=20$ samples which can approximate the non-Markovian environment with a high precision \cite{sup}.}

(ii) {For the target system $H_\text{S}= \frac{\omega_0 }{2}\sum_{i=1}^n\sigma^z_{i}$ with $\omega_0$ being the energy splitting to be measured, let the probes precess unperturbedly for a duration $t$ under the non-Markovian noise. This non-Markovian circumstance is simulated by the aforementioned bath-engineering technique. Explicitly, we introduce an additional time-dependent Hamiltonian term in the form of $H_\text{SB}^m(t)=\sum_{i=1}^n\beta^m_{i}(t)\sigma^z_{i}$, where $\beta_{i}^m(t)$'s are stochastic errors subject to Drude-Lorentz spectral density as shown in Eq.~(\ref{eq:beta}), and $m$ is the sampling number so as to approximate the non-Markovian noise via the ensemble-averaging approach \cite{sup}. So the total Hamiltonian in the $m$-th experiment can be written as $H_\text{S}+H_\text{SB}^m(t)$. Note that the evolution of this Hamiltonian corresponds to a collective single-qubit rotation about the $z$-axis, which can be easily implemented in most physical systems. In addition, this time-dependent Hamiltonian can be discretized into time-independent slices for experimental realization (here we use $10^3$ slices \cite{sup}). Regarding the experimental parameters, we set $\omega_0=10$~kHz and choose $m=20$ samples for ensemble average, which can already approximate the non-Markovian environment with a high precision \cite{sup}.}

\begin{figure*}
\includegraphics[scale=0.8]{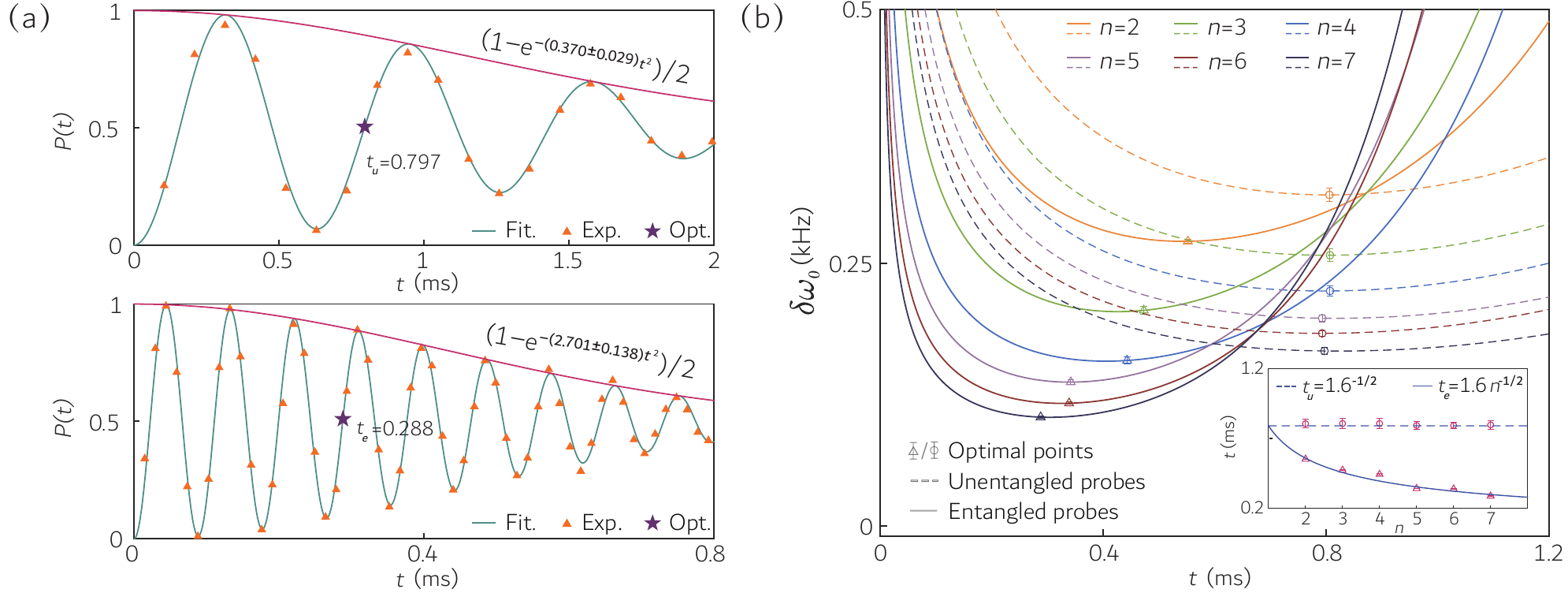}
\caption{(a) {Dynamics of population $P(t)$ for the unentangled (top panel) and entangled (bottom panel) probes.} Blue curves are obtained by numerically fitting the experimental data (yellow triangles) using Eq.~(\ref{population}), and red lines are envelopes with the form $(1-e^{-\Gamma(t)})/2$, which depicts the time-dependent decoherence factor $\Gamma(t)$. The star symbol represents the optimal measurement point determined by the experiment. (b) Standard deviation $\delta\omega_0$ of the measured frequency with respect to the variation of $t$. {Dashed (solid) curves are fitting results corresponding to the unentangled (entangled) probes for different number of qubits. Triangles and dots are the optimal measurement points. The scaling behavior of these optimal times when changing $n$ is shown in the inset. }}
\label{fig:7qubit}
\end{figure*}
%and the phases accumulated at time $t$ is $\theta_m^j(t)=[\frac{\omega_0 }{2}+\beta^z_{j}(m)]t$; see Fig.~\ref{fig:circuit}(b). The evolution of $H_m(t)$ corresponds to collective single-qubit rotations about the $z$-axis, which can be realized by a concatenation of rotations in the transverse plane via $R_z(\theta) = R_x(\frac{\pi}{2})R_y(\theta) R_x(-\frac{\pi}{2})$. We optimize a 2-ms-shaped pulse to engineer its dynamics for a duration $t$. Numerical simulations show that this ensemble average over 20 samples can approximate the non-Markovian environment with a high precision \cite{sup}. We repeat the evolution $H_m(t)$ for $M=20$ times with different noise distributions $\beta^z_{j}(t)$ and discretize the time-dependent $H_m(t)$ into $K=10^3$ slices with time-independent Hamiltonians $H_m(k)$ \cite{sup}.

%(iii) Read out the accumulated phase. Unlike the standard Ramsey interferometry that maps the phase onto a population difference, the phase can be extracted more straightforwardly using NMR techniques. For the unentangled probes, the phase is encoded in each individual qubit, so the readout only involves single-coherence measurement, which is standard in NMR spectroscopy. For the entangled probes, since the final phase is retained by the highest coherence, we apply the multi-coherence measurement (MCM) \cite{Laflamme2002a} approach for its precise determination; see Fig.~\ref{fig:circuit}(b). Subsequently, the transition probabilities are computed by averaging over the ensemble with $M=20$ samples to simulate the non-Markovian decay.

(iii) {Read out the accumulated phase. For the unentangled probes, the final phase is encoded in each individual qubit, so the readout only involves single-qubit measurements, which is similar to the standard Ramsey interferometry experiment. For the entangled probes, the final phase is retained in the relative phase between $|00...0\rangle$ and $|11...1\rangle$, and we apply the multi-coherence measurement (MCM) technique \cite{Laflamme2002a} to extract that phase. As shown in Fig.~\ref{fig:circuit}(b), this technique involves a series of elementary quantum gates, which can be extended to other systems as well. Subsequently, the transition probabilities are computed by averaging over the ensemble with $m=20$ samples to simulate the non-Markovian decay.}

\emph{Results.}---{We have conducted Ramsey experiments for $n=2\sim7$ qubits. Taking the 7-qubit case as an example, the transition probability $P(t)$ is shown in Fig.~\ref{fig:7qubit}(a) for both the unentangled (top panel) and entangled (bottom panel) probes.} Compared to the unentangled case, $P(t)$ in the entangled case manifests a much faster oscillation owing to the rapid accumulation ($\sim 7$ times) of phase. We use the theoretical form
\begin{equation}
P(t)=\frac{1}{2}[1-\cos(n\omega_0 t)e^{-n\Gamma(t)}]
\label{population}
\end{equation}
to fit the experimental data which gives the energy-splitting frequency $\omega_0$ and decoherence factor $\Gamma(t)$. For the unentangled probes, the measured frequency is $\omega_0|_u=9.871$~kHz and the decoherence factor is $\Gamma_u = (0.370\pm0.029)t^2$. For the entangled probes, the measured frequency is $\omega_0|_e=10.142$~kHz and the decoherence factor is $\Gamma_e=(2.701\pm0.138)t^2$.
The minimum standard deviations of the frequencies are $\delta\omega_0|_u= 0.169\pm0.003$~kHz and $\delta\omega_0|_e= 0.105\pm0.001$~kHz at the optimal measurement times $t_u=0.797\pm0.032$~ms and $t_e=0.288\pm0.003$~ms, respectively. Hence, the ratio of the sensitivity enhancement is $r=\delta\omega_0|_u/\delta\omega_0|_e = 1.608\pm0.035\approx 7^{\frac{1}{4}}$, indicating that the entangled probes are indeed superior to the unentangled ones under the non-Markovian environment. {Note that the readout of the NMR experiment is from the ensemble average result, which means the statistical error by repeating experiments is almost negligible. So, in this work, the error bars represent experimental uncertainties by error propagation from the fitting uncertainties in Fig.~\ref{fig:7qubit}(a);  see SM for details \cite{sup}.}

\begin{figure}
\includegraphics[scale=1]{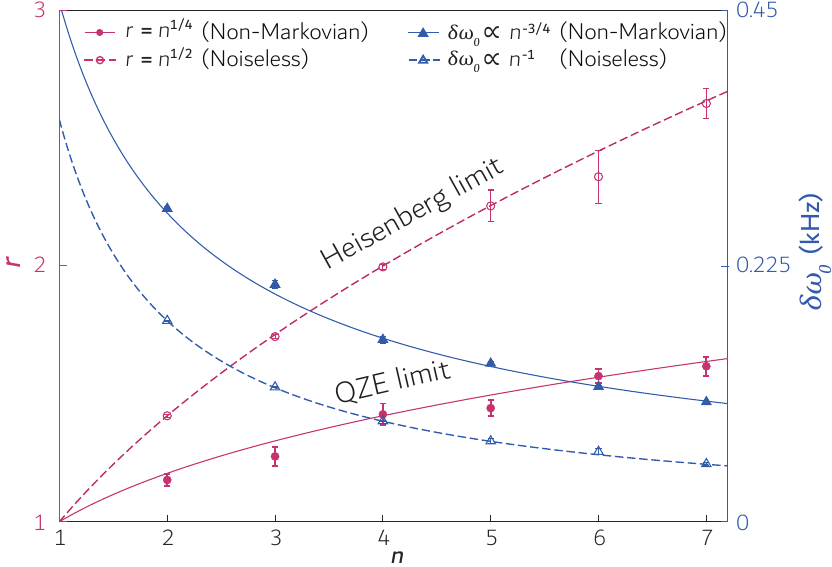}\caption{Experimental results of the ratio $r = \delta\omega_0|_u/\delta\omega_0|_e$ (dots) and the minimum standard deviation $\delta\omega_0|_e$ (triangles) for different number of qubits $n$. Under the non-Markovian noise, the sensitivity is enhanced by $r = n^{1/4}$ (solid red) with the entangled probes, reaching the QZE limit; $\delta\omega_0|_e$ scales as $n ^{-3/4}$ (solid  blue). In the absence of noise, the sensitivity enhancement is governed by the Heisenberg limit $r = n^{1/2}$ (dashed red), and $\delta\omega_0|_e$ scales as $n ^{-1}$ (dashed blue).}
\label{fig:r}
\end{figure}

For the other numbers of qubits, the experimentally measured frequencies $\omega_u$ and $\omega_e$, in association with their respective decoherence factors $\Gamma_u$ and $\Gamma_e$, can be found in SM \cite{sup}. The standard deviation of the estimated frequency is obtained by processing the experimental data, as shown in Fig.~\ref{fig:7qubit}(b), {where the dashed and solid lines are for the
unentangled and entangled probes,} respectively. For the unentangled probes, the optimal measurement time $t_u$ is almost the same for different $n$'s and the minimum standard deviation $\delta\omega_0|_u$ is proportional to $n^{-1/2}$. For the entangled probes, $t_e$ and $\delta\omega_0|_e$ decrease with the growth of $n$. By varying the number of qubits, we compute the ratio $r = \delta\omega_0|_u/\delta\omega_0|_e$ to estimate the improvement of sensitivity with the entangled probes; see the solid red curve in Fig.~\ref{fig:r}. As predicted, the experimentally-observed $r$ scales as $r \sim n^{1/4}$, which is the QZE limit. For comparison, we have performed another group of experiments in the absence of noise \cite{sup}. In this noiseless scenario, the ratio of sensitivity $r$ reaches the Heisenberg limit, scaling as $r \sim n^{1/2}$ by the dashed red curve in Fig.~\ref{fig:r}.

In addition, we depict distinct tendencies of the minimum standard deviation $\delta\omega_0|_e$ of the entangled probes for non-Markovian and noiseless environments, respectively. As shown in Fig.~\ref{fig:r}, with the growth in number of qubits, experimental results show that $\delta\omega_0|_e \propto n ^{-3/4}$ for the non-Markovian noise by solid blue curve, while $\delta\omega_0|_e \propto n ^{-1}$ for the noiseless case by dashed blue curve. Definitely, our high-precision experiment is sufficient to verify that the measurement accuracy of the entangled probes can reach the QZE limit under the non-Markovian noise.

%\section*{Conclusion}
\textit{Conclusion.}---We experimentally demonstrate that the entangled probes can enhance the sensitivity of quantum metrology by the QZE. When the coherence decays quadratically with time, the entangled probes optimize the time for performing the measurement. On the contrary, when the decoherence rate is time-independent, the entangled probes cannot effectively improve the precision but shortens the measurement time. We remark that since our quantum simulation approach can subtly engineer the parameters of both the system and bath, it may provide an effective platform for experimentally verifying various quantum metrology schemes, e.g., achieving ideal precision by coupling to a bath with a band structure~\cite{Bai2019a}. And it can also be implemented in other physical systems \cite{Soare2014a,Potocnik2018}.
%
%2. We also show that not the precision but the measurement time can
%be improved by using entangled probes in the Markovian bath.
%
%3. Possible future work?

%\section*{Acknowledgement}
\textit{Acknowledgement.}---This work is supported by the National Key Research and Development Program of China (2019YFA0308100), Beijing Natural
Science Foundation (1202017), Beijing Normal University (2022129), National Natural Science Foundation of China (11674033, 12075110, 11975117, 11905099, 11875159, 11905111 and U1801661),  Guangdong Basic and Applied Basic Research Foundation (2019A1515011383), Guangdong International Collaboration Program (2020A0505100001), Science, Technology and Innovation Commission of Shenzhen Municipality (ZDSYS20190902092905285, KQTD20190929173815000, JCYJ20200109140803865,  and JCYJ20180302174036418),  Pengcheng Scholars, Guangdong Innovative and Entrepreneurial Research Team Program (2019ZT08C044), and Guangdong Provincial Key Laboratory (2019B121203002).

%\bibliography{aa}

%\bibliography{reference2}

\begin{thebibliography}{37}%
\makeatletter
\providecommand \@ifxundefined [1]{%
 \@ifx{#1\undefined}
}%
\providecommand \@ifnum [1]{%
 \ifnum #1\expandafter \@firstoftwo
 \else \expandafter \@secondoftwo
 \fi
}%
\providecommand \@ifx [1]{%
 \ifx #1\expandafter \@firstoftwo
 \else \expandafter \@secondoftwo
 \fi
}%
\providecommand \natexlab [1]{#1}%
\providecommand \enquote  [1]{``#1''}%
\providecommand \bibnamefont  [1]{#1}%
\providecommand \bibfnamefont [1]{#1}%
\providecommand \citenamefont [1]{#1}%
\providecommand \href@noop [0]{\@secondoftwo}%
\providecommand \href [0]{\begingroup \@sanitize@url \@href}%
\providecommand \@href[1]{\@@startlink{#1}\@@href}%
\providecommand \@@href[1]{\endgroup#1\@@endlink}%
\providecommand \@sanitize@url [0]{\catcode `\\12\catcode `\$12\catcode
  `\&12\catcode `\#12\catcode `\^12\catcode `\_12\catcode `\%12\relax}%
\providecommand \@@startlink[1]{}%
\providecommand \@@endlink[0]{}%
\providecommand \url  [0]{\begingroup\@sanitize@url \@url }%
\providecommand \@url [1]{\endgroup\@href {#1}{\urlprefix }}%
\providecommand \urlprefix  [0]{URL }%
\providecommand \Eprint [0]{\href }%
\providecommand \doibase [0]{http://dx.doi.org/}%
\providecommand \selectlanguage [0]{\@gobble}%
\providecommand \bibinfo  [0]{\@secondoftwo}%
\providecommand \bibfield  [0]{\@secondoftwo}%
\providecommand \translation [1]{[#1]}%
\providecommand \BibitemOpen [0]{}%
\providecommand \bibitemStop [0]{}%
\providecommand \bibitemNoStop [0]{.\EOS\space}%
\providecommand \EOS [0]{\spacefactor3000\relax}%
\providecommand \BibitemShut  [1]{\csname bibitem#1\endcsname}%
\let\auto@bib@innerbib\@empty
%</preamble>
\bibitem [{\citenamefont {Caves}(1981)}]{Caves1981}%
  \BibitemOpen
  \bibfield  {author} {\bibinfo {author} {\bibfnamefont {C.~M.}\ \bibnamefont
  {Caves}},\ }\bibfield  {title} {\bibinfo {title} {Quantum-mechanical noise in
  an interferometer},\ }\href {\doibase 10.1103/PhysRevD.23.1693} {\bibfield  {journal}
  {\bibinfo  {journal} {Phys. Rev. D}\ }\textbf {\bibinfo {volume} {23}},\
  \bibinfo {pages} {1693} (\bibinfo {year} {1981})}\BibitemShut {NoStop}%
\bibitem [{\citenamefont {Giovannetti}\ \emph {et~al.}(2004)\citenamefont
  {Giovannetti}, \citenamefont {Lloyd},\ and\ \citenamefont
  {Maccone}}]{Giovannetti2004a}%
  \BibitemOpen
  \bibfield  {author} {\bibinfo {author} {\bibfnamefont {V.}~\bibnamefont
  {Giovannetti}}, \bibinfo {author} {\bibfnamefont {S.}~\bibnamefont {Lloyd}},
  \ and\ \bibinfo {author} {\bibfnamefont {L.}~\bibnamefont {Maccone}},\ }\bibfield  {title} {\bibinfo {title} {Quantum-enhanced measurements: Beating the standard quantum limit},\ }\href
  {\doibase 10.1126/science.1104149} {\bibfield  {journal} {\bibinfo  {journal}
  {Science}\ }\textbf {\bibinfo {volume} {306}},\ \bibinfo {pages} {1330}
  (\bibinfo {year} {2004})}\BibitemShut {NoStop}%
\bibitem [{\citenamefont {Pezz{\`e}}\ \emph {et~al.}(2018)\citenamefont
  {Pezz{\`e}}, \citenamefont {Smerzi}, \citenamefont {Oberthaler},
  \citenamefont {Schmied},\ and\ \citenamefont {Treutlein}}]{Pezze2018}%
  \BibitemOpen
  \bibfield  {author} {\bibinfo {author} {\bibfnamefont {L.}~\bibnamefont
  {Pezz{\`e}}}, \bibinfo {author} {\bibfnamefont {A.}~\bibnamefont {Smerzi}},
  \bibinfo {author} {\bibfnamefont {M.~K.}\ \bibnamefont {Oberthaler}},
  \bibinfo {author} {\bibfnamefont {R.}~\bibnamefont {Schmied}}, \ and\
  \bibinfo {author} {\bibfnamefont {P.}~\bibnamefont {Treutlein}},\ }\bibfield  {title}
  {\bibinfo {title} {Quantum metrology with nonclassical states of atomic ensembles},\ }\href
  {\doibase 10.1103/RevModPhys.90.035005} {\bibfield  {journal} {\bibinfo
  {journal} {Rev. Mod. Phys.}\ }\textbf {\bibinfo {volume} {90}},\ \bibinfo
  {pages} {035005} (\bibinfo {year} {2018})}\BibitemShut {NoStop}%
\bibitem [{\citenamefont {Breuer}\ and\ \citenamefont
  {Petruccione}(2002)}]{Breuer2002}%
  \BibitemOpen
  \bibfield  {author} {\bibinfo {author} {\bibfnamefont {H.~P.}\ \bibnamefont
  {Breuer}}\ and\ \bibinfo {author} {\bibfnamefont {F.}~\bibnamefont
  {Petruccione}},\ }\href@noop {} {\emph {\bibinfo {title} {The Theory of Open
  Quantum Systems}}}\ (\bibinfo  {publisher} {Oxford University Presss},\
  \bibinfo {year} {2002})\BibitemShut {NoStop}%
\bibitem [{\citenamefont {Huelga}\ \emph {et~al.}(1997)\citenamefont {Huelga},
  \citenamefont {Macchiavello}, \citenamefont {Pellizzari}, \citenamefont
  {Ekert}, \citenamefont {Plenio},\ and\ \citenamefont {Cirac}}]{Huelga1997}%
  \BibitemOpen
  \bibfield  {author} {\bibinfo {author} {\bibfnamefont {S.~F.}\ \bibnamefont
  {Huelga}}, \bibinfo {author} {\bibfnamefont {C.}~\bibnamefont
  {Macchiavello}}, \bibinfo {author} {\bibfnamefont {T.}~\bibnamefont
  {Pellizzari}}, \bibinfo {author} {\bibfnamefont {A.~K.}\ \bibnamefont
  {Ekert}}, \bibinfo {author} {\bibfnamefont {M.~B.}\ \bibnamefont {Plenio}}, \
  and\ \bibinfo {author} {\bibfnamefont {J.~I.}\ \bibnamefont {Cirac}},\ }\bibfield  {title} {\bibinfo {title} {Improvement of frequency standards with quantum entanglement},\ }\href
  {\doibase 10.1103/PhysRevLett.79.3865} {\bibfield  {journal} {\bibinfo
  {journal} {Phys. Rev. Lett.}\ }\textbf {\bibinfo {volume} {79}},\ \bibinfo
  {pages} {3865} (\bibinfo {year} {1997})}\BibitemShut {NoStop}%
\bibitem [{\citenamefont {Ma}\ \emph {et~al.}(2011)\citenamefont {Ma},
  \citenamefont {Wang}, \citenamefont {Sun},\ and\ \citenamefont
  {Nori}}]{Ma2011b}%
  \BibitemOpen
  \bibfield  {author} {\bibinfo {author} {\bibfnamefont {J.}~\bibnamefont
  {Ma}}, \bibinfo {author} {\bibfnamefont {X.}~\bibnamefont {Wang}}, \bibinfo
  {author} {\bibfnamefont {C.~P.}\ \bibnamefont {Sun}}, \ and\ \bibinfo
  {author} {\bibfnamefont {F.}~\bibnamefont {Nori}},\ }\bibfield  {title} {\bibinfo
  {title} {Quantum spin squeezing},\ }\href {\doibase
  10.1016/j.physrep.2011.08.003} {\bibfield  {journal} {\bibinfo  {journal}
  {Phys. Rep.}\ }\textbf {\bibinfo {volume} {509}},\ \bibinfo {pages} {89}
  (\bibinfo {year} {2011})}\BibitemShut {NoStop}%
\bibitem [{\citenamefont {Yamamoto}\ \emph {et~al.}(2022)\citenamefont
  {Yamamoto}, \citenamefont {Endo}, \citenamefont {Hakoshima}, \citenamefont
  {Matsuzaki},\ and\ \citenamefont {Tokunaga}}]{Yamamoto2022}%
  \BibitemOpen
  {\bibfield  {author} {\bibinfo {author} {\bibfnamefont {K.}~\bibnamefont
  {Yamamoto}}, \bibinfo {author} {\bibfnamefont {S.}~\bibnamefont {Endo}},
  \bibinfo {author} {\bibfnamefont {H.}~\bibnamefont {Hakoshima}}, \bibinfo
  {author} {\bibfnamefont {Y.}~\bibnamefont {Matsuzaki}}, \ and\ \bibinfo
  {author} {\bibfnamefont {Y.}~\bibnamefont {Tokunaga}},\ }\bibfield  {title} {\bibinfo {title} {Error-mitigated quantum metrology via virtual purification},\ }\href {\doibase
  10.48550/arXiv.2112.01850} {\bibfield  {journal} {\bibinfo  {journal}
  {arXiv:2112.01850}\ } (\bibinfo {year} {2022})}}\BibitemShut {NoStop}%
\bibitem [{\citenamefont {Matsuzaki}\ \emph {et~al.}(2018)\citenamefont
  {Matsuzaki}, \citenamefont {Benjamin}, \citenamefont {Nakayama},
  \citenamefont {Saito},\ and\ \citenamefont {Munro}}]{Matsuzaki2018}%
  \BibitemOpen
  {\bibfield  {author} {\bibinfo {author} {\bibfnamefont {Y.}~\bibnamefont
  {Matsuzaki}}, \bibinfo {author} {\bibfnamefont {S.}~\bibnamefont {Benjamin}},
  \bibinfo {author} {\bibfnamefont {S.}~\bibnamefont {Nakayama}}, \bibinfo
  {author} {\bibfnamefont {S.}~\bibnamefont {Saito}}, \ and\ \bibinfo {author}
  {\bibfnamefont {W.~J.}\ \bibnamefont {Munro}},\ }\bibfield  {title} {\bibinfo {title} {Quantum metrology beyond the classical limit under the effect of dephasing},\ }\href {\doibase
  10.1103/PhysRevLett.120.140501} {\bibfield  {journal} {\bibinfo  {journal}
  {Phys. Rev. Lett.}\ }\textbf {\bibinfo {volume} {120}},\ \bibinfo {pages}
  {140501} (\bibinfo {year} {2018})}}\BibitemShut {NoStop}%
\bibitem [{\citenamefont {Bai}\ \emph {et~al.}(2019)\citenamefont {Bai},
  \citenamefont {Peng}, \citenamefont {Luo},\ and\ \citenamefont
  {An}}]{Bai2019a}%
  \BibitemOpen
  \bibfield  {author} {\bibinfo {author} {\bibfnamefont {K.}~\bibnamefont
  {Bai}}, \bibinfo {author} {\bibfnamefont {Z.}~\bibnamefont {Peng}}, \bibinfo
  {author} {\bibfnamefont {H.-G.}\ \bibnamefont {Luo}}, \ and\ \bibinfo
  {author} {\bibfnamefont {J.-H.}\ \bibnamefont {An}},\ }\bibfield  {title} {\bibinfo
  {title} {Retrieving ideal precision in noisy quantum optical metrology},\ }\href {\doibase
  10.1103/PhysRevLett.123.040402} {\bibfield  {journal} {\bibinfo  {journal}
  {Phys. Rev. Lett.}\ }\textbf {\bibinfo {volume} {123}},\ \bibinfo {pages}
  {040402} (\bibinfo {year} {2019})}\BibitemShut {NoStop}%
\bibitem [{\citenamefont {Ai}\ and\ \citenamefont {Liao}(2010)}]{Ai2010}%
  \BibitemOpen
  \bibfield  {author} {\bibinfo {author} {\bibfnamefont {Q.}~\bibnamefont
  {Ai}}\ and\ \bibinfo {author} {\bibfnamefont {J.-Q.}\ \bibnamefont {Liao}},\
  }\bibfield  {title} {\bibinfo {title} {{Quantum anti-Zeno effect in artificial quantum systems}},\ }\href {\doibase 10.1088/0253-6102/54/6/07} {\bibfield  {journal} {\bibinfo
  {journal} {Commun. Theor. Phys.}\ }\textbf {\bibinfo {volume} {54}},\
  \bibinfo {pages} {985} (\bibinfo {year} {2010})}\BibitemShut {NoStop}%
\bibitem [{\citenamefont {Monz}\ \emph {et~al.}(2011)\citenamefont {Monz},
  \citenamefont {Schindler}, \citenamefont {Barreiro}, \citenamefont {Chwalla},
  \citenamefont {Nigg}, \citenamefont {Coish}, \citenamefont {Harlander},
  \citenamefont {H{\"a}nsel}, \citenamefont {Hennrich},\ and\ \citenamefont
  {Blatt}}]{Monz2011}%
  \BibitemOpen
  \bibfield  {author} {\bibinfo {author} {\bibfnamefont {T.}~\bibnamefont
  {Monz}}, \bibinfo {author} {\bibfnamefont {P.}~\bibnamefont {Schindler}},
  \bibinfo {author} {\bibfnamefont {J.~T.}\ \bibnamefont {Barreiro}}, \bibinfo
  {author} {\bibfnamefont {M.}~\bibnamefont {Chwalla}}, \bibinfo {author}
  {\bibfnamefont {D.}~\bibnamefont {Nigg}}, \bibinfo {author} {\bibfnamefont
  {W.~A.}\ \bibnamefont {Coish}}, \bibinfo {author} {\bibfnamefont
  {M.}~\bibnamefont {Harlander}}, \bibinfo {author} {\bibfnamefont
  {W.}~\bibnamefont {H{\"a}nsel}}, \bibinfo {author} {\bibfnamefont
  {M.}~\bibnamefont {Hennrich}}, \ and\ \bibinfo {author} {\bibfnamefont
  {R.}~\bibnamefont {Blatt}},\ }\bibfield  {title} {\bibinfo {title} {14-qubit
  entanglement: Creation and coherence},\ }\href {\doibase 10.1103/PhysRevLett.106.130506}
  {\bibfield  {journal} {\bibinfo  {journal} {Phys. Rev. Lett.}\ }\textbf
  {\bibinfo {volume} {106}},\ \bibinfo {pages} {130506} (\bibinfo {year}
  {2011})}\BibitemShut {NoStop}%
\bibitem [{\citenamefont {He}\ \emph {et~al.}(2021)\citenamefont {He},
  \citenamefont {Guang}, \citenamefont {Li}, \citenamefont {Deng},
  \citenamefont {Zhang}, \citenamefont {Zhao}, \citenamefont {Deng},\ and\
  \citenamefont {Ai}}]{He2021}%
  \BibitemOpen
  \bibfield  {author} {\bibinfo {author} {\bibfnamefont {W.-T.}\ \bibnamefont
  {He}}, \bibinfo {author} {\bibfnamefont {H.-Y.}\ \bibnamefont {Guang}},
  \bibinfo {author} {\bibfnamefont {Z.-Y.}\ \bibnamefont {Li}}, \bibinfo
  {author} {\bibfnamefont {R.-Q.}\ \bibnamefont {Deng}}, \bibinfo {author}
  {\bibfnamefont {N.-N.}\ \bibnamefont {Zhang}}, \bibinfo {author}
  {\bibfnamefont {J.-X.}\ \bibnamefont {Zhao}}, \bibinfo {author}
  {\bibfnamefont {F.-G.}\ \bibnamefont {Deng}}, \ and\ \bibinfo {author}
  {\bibfnamefont {Q.}~\bibnamefont {Ai}},\ }\bibfield  {title} {\bibinfo
  {title} {Quantum metrology with one auxiliary particle in a correlated bath
  and its quantum simulation},\ }\href {\doibase
  10.1103/PhysRevA.104.062429} {\bibfield  {journal} {\bibinfo  {journal}
  {Phys. Rev. A}\ }\textbf {\bibinfo {volume} {104}},\ \bibinfo {pages}
  {062429} (\bibinfo {year} {2021})}\BibitemShut {NoStop}%
\bibitem [{\citenamefont {Kukita}\ \emph {et~al.}(2021)\citenamefont {Kukita},
  \citenamefont {Matsuzaki},\ and\ \citenamefont {Kondo}}]{Kukita2021}%
  \BibitemOpen
  {\bibfield  {author} {\bibinfo {author} {\bibfnamefont {S.}~\bibnamefont
  {Kukita}}, \bibinfo {author} {\bibfnamefont {Y.}~\bibnamefont {Matsuzaki}}, \
  and\ \bibinfo {author} {\bibfnamefont {Y.}~\bibnamefont {Kondo}},\ }\bibfield  {title} {\bibinfo {title} {Heisenberg-limited quantum metrology using collective dephasing},\ }\href
  {\doibase 10.1103/PhysRevApplied.16.064026} {\bibfield  {journal} {\bibinfo
  {journal} {Phys. Rev. Applied}\ }\textbf {\bibinfo {volume} {16}},\ \bibinfo
  {pages} {064026} (\bibinfo {year} {2021})}}\BibitemShut {NoStop}%
\bibitem [{\citenamefont {Boixo}\ \emph {et~al.}(2007)\citenamefont {Boixo},
  \citenamefont {Flammia}, \citenamefont {Caves},\ and\ \citenamefont
  {Geremia}}]{Boixo2007}%
  \BibitemOpen
  \bibfield  {author} {\bibinfo {author} {\bibfnamefont {S.}~\bibnamefont
  {Boixo}}, \bibinfo {author} {\bibfnamefont {S.~T.}\ \bibnamefont {Flammia}},
  \bibinfo {author} {\bibfnamefont {C.~M.}\ \bibnamefont {Caves}}, \ and\
  \bibinfo {author} {\bibfnamefont {J.~M.}~\bibnamefont {Geremia}},\ }\bibfield
  {title} {\bibinfo {title} {Generalized limits for single-parameter quantum
  estimation},\ }\href
  {\doibase 10.1103/PhysRevLett.98.090401} {\bibfield  {journal} {\bibinfo
  {journal} {Phys. Rev. Lett.}\ }\textbf {\bibinfo {volume} {98}},\ \bibinfo
  {pages} {090401} (\bibinfo {year} {2007})}\BibitemShut {NoStop}%
\bibitem [{\citenamefont {Nie}\ \emph {et~al.}(2015)\citenamefont {Nie},
  \citenamefont {Li}, \citenamefont {Cui}, \citenamefont {Luo}, \citenamefont
  {Huang}, \citenamefont {Chen}, \citenamefont {Lee}, \citenamefont {Peng},\
  and\ \citenamefont {Du}}]{Nie2015}%
  \BibitemOpen
  \bibfield  {author} {\bibinfo {author} {\bibfnamefont {X.}~\bibnamefont
  {Nie}}, \bibinfo {author} {\bibfnamefont {J.}~\bibnamefont {Li}}, \bibinfo
  {author} {\bibfnamefont {J.}~\bibnamefont {Cui}}, \bibinfo {author}
  {\bibfnamefont {Z.}~\bibnamefont {Luo}}, \bibinfo {author} {\bibfnamefont
  {J.}~\bibnamefont {Huang}}, \bibinfo {author} {\bibfnamefont
  {H.}~\bibnamefont {Chen}}, \bibinfo {author} {\bibfnamefont {C.}~\bibnamefont
  {Lee}}, \bibinfo {author} {\bibfnamefont {X.}~\bibnamefont {Peng}}, \ and\
  \bibinfo {author} {\bibfnamefont {J.}~\bibnamefont {Du}},\ }\bibfield  {title} {\bibinfo {title} {Quantum simulation of interaction blockade in a two-site {{Bose}}\textendash{{Hubbard}} system with solid quadrupolar crystal},\
  }\href {\doibase
  10.1088/1367-2630/17/5/053028} {\bibfield  {journal} {\bibinfo  {journal}
  {New J. Phys.}\ }\textbf {\bibinfo {volume} {17}},\ \bibinfo
  {pages} {053028} (\bibinfo {year} {2015})}\BibitemShut {NoStop}%
\bibitem [{\citenamefont {Choi}\ and\ \citenamefont
  {Sundaram}(2008)}]{Choi2008}%
  \BibitemOpen
  \bibfield  {author} {\bibinfo {author} {\bibfnamefont {S.}~\bibnamefont
  {Choi}}\ and\ \bibinfo {author} {\bibfnamefont {B.}~\bibnamefont
  {Sundaram}},\ }\bibfield  {title} {\bibinfo {title} {Bose-Einstein condensate
  as a nonlinear Ramsey interferometer operating beyond the Heisenberg limit},\
  }\href {\doibase 10.1103/PhysRevA.77.053613} {\bibfield
  {journal} {\bibinfo  {journal} {Phys. Rev. A}\ }\textbf {\bibinfo {volume}
  {77}},\ \bibinfo {pages} {053613} (\bibinfo {year} {2008})}\BibitemShut
  {NoStop}%
\bibitem [{\citenamefont {Roy}\ and\ \citenamefont
  {Braunstein}(2008)}]{Roy2008}%
  \BibitemOpen
  \bibfield  {author} {\bibinfo {author} {\bibfnamefont {S.~M.}\ \bibnamefont
  {Roy}}\ and\ \bibinfo {author} {\bibfnamefont {S.~L.}\ \bibnamefont
  {Braunstein}},\ }\bibfield  {title} {\bibinfo {title} {Exponentially enhanced
  quantum metrology},\ }\href {\doibase 10.1103/PhysRevLett.100.220501} {\bibfield
  {journal} {\bibinfo  {journal} {Phys. Rev. Lett.}\ }\textbf {\bibinfo
  {volume} {100}},\ \bibinfo {pages} {220501} (\bibinfo {year}
  {2008})}\BibitemShut {NoStop}%
\bibitem [{\citenamefont {Napolitano}\ \emph {et~al.}(2011)\citenamefont
  {Napolitano}, \citenamefont {Koschorreck}, \citenamefont {Dubost},
  \citenamefont {Behbood}, \citenamefont {Sewell},\ and\ \citenamefont
  {Mitchell}}]{Napolitano2011}%
  \BibitemOpen
  \bibfield  {author} {\bibinfo {author} {\bibfnamefont {M.}~\bibnamefont
  {Napolitano}}, \bibinfo {author} {\bibfnamefont {M.}~\bibnamefont
  {Koschorreck}}, \bibinfo {author} {\bibfnamefont {B.}~\bibnamefont {Dubost}},
  \bibinfo {author} {\bibfnamefont {N.}~\bibnamefont {Behbood}}, \bibinfo
  {author} {\bibfnamefont {R.~J.}\ \bibnamefont {Sewell}}, \ and\ \bibinfo
  {author} {\bibfnamefont {M.~W.}\ \bibnamefont {Mitchell}},\ }\bibfield
  {title} {\bibinfo {title} {{Interaction-based quantum metrology showing
  scaling beyond the Heisenberg limit}},\ }\href {\doibase
  10.1038/nature09778} {\bibfield  {journal} {\bibinfo  {journal} {Nature}\
  }\textbf {\bibinfo {volume} {471}},\ \bibinfo {pages} {486} (\bibinfo {year}
  {2011})}\
  \BibitemShut {NoStop}%
\bibitem [{\citenamefont {Nakazato}\ \emph {et~al.}(1996)\citenamefont
  {Nakazato}, \citenamefont {Namiki},\ and\ \citenamefont
  {Pascazio}}]{Nakazato1996}%
  \BibitemOpen
  \bibfield  {author} {\bibinfo {author} {\bibfnamefont {H.}~\bibnamefont
  {Nakazato}}, \bibinfo {author} {\bibfnamefont {M.}~\bibnamefont {Namiki}}, \
  and\ \bibinfo {author} {\bibfnamefont {S.}~\bibnamefont {Pascazio}},\ }\bibfield  {title} {\bibinfo {title} {Temporal behavior of quantum mechanical systems},\ }\href
  {\doibase 10.1142/S0217979296000118} {\bibfield  {journal} {\bibinfo
  {journal} {Int. J. Mod. Phys. B}\ }\textbf {\bibinfo {volume} {10}},\
  \bibinfo {pages} {247} (\bibinfo {year} {1996})}\BibitemShut {NoStop}%
\bibitem [{\citenamefont {Ai}\ \emph {et~al.}(2010)\citenamefont {Ai},
  \citenamefont {Li}, \citenamefont {Zheng},\ and\ \citenamefont
  {Sun}}]{Ai2010a}%
  \BibitemOpen
  \bibfield  {author} {\bibinfo {author} {\bibfnamefont {Q.}~\bibnamefont
  {Ai}}, \bibinfo {author} {\bibfnamefont {Y.}~\bibnamefont {Li}}, \bibinfo
  {author} {\bibfnamefont {H.}~\bibnamefont {Zheng}}, \ and\ \bibinfo {author}
  {\bibfnamefont {C.~P.}\ \bibnamefont {Sun}},\ }\bibfield  {title} {\bibinfo
  {title} {Quantum anti-Zeno effect without rotating wave approximation},\
  }\href {\doibase
  10.1103/PhysRevA.81.042116} {\bibfield  {journal} {\bibinfo  {journal} {Phys.
  Rev. A}\ }\textbf {\bibinfo {volume} {81}},\ \bibinfo {pages} {042116}
  (\bibinfo {year} {2010})}\BibitemShut {NoStop}%
\bibitem [{\citenamefont {Ai}\ \emph {et~al.}(2013)\citenamefont {Ai},
  \citenamefont {Xu}, \citenamefont {Yi}, \citenamefont {Kofman}, \citenamefont
  {Sun},\ and\ \citenamefont {Nori}}]{Ai2013}%
  \BibitemOpen
  \bibfield  {author} {\bibinfo {author} {\bibfnamefont {Q.}~\bibnamefont
  {Ai}}, \bibinfo {author} {\bibfnamefont {D.}~\bibnamefont {Xu}}, \bibinfo
  {author} {\bibfnamefont {S.}~\bibnamefont {Yi}}, \bibinfo {author}
  {\bibfnamefont {A.~G.}\ \bibnamefont {Kofman}}, \bibinfo {author}
  {\bibfnamefont {C.~P.}\ \bibnamefont {Sun}}, \ and\ \bibinfo {author}
  {\bibfnamefont {F.}~\bibnamefont {Nori}},\ }\bibfield  {title} {\bibinfo
  {title} {Quantum anti-Zeno effect without wave function reduction},\
  }\href {\doibase
  10.1038/srep01752} {\bibfield  {journal} {\bibinfo  {journal} {Sci Rep}\
  }\textbf {\bibinfo {volume} {3}},\ \bibinfo {pages} {1752} (\bibinfo {year}
  {2013})}\BibitemShut {NoStop}%
\bibitem [{\citenamefont {Harrington}\ \emph {et~al.}(2017)\citenamefont
  {Harrington}, \citenamefont {Monroe},\ and\ \citenamefont
  {Murch}}]{Harrington2017}%
  \BibitemOpen
  \bibfield  {author} {\bibinfo {author} {\bibfnamefont {P.~M.}\ \bibnamefont
  {Harrington}}, \bibinfo {author} {\bibfnamefont {J.~T.}\ \bibnamefont
  {Monroe}}, \ and\ \bibinfo {author} {\bibfnamefont {K.~W.}\ \bibnamefont
  {Murch}},\ }\bibfield  {title} {\bibinfo {title} {Quantum Zeno effects from
  measurement controlled qubit-bath interactions},\ }\href {\doibase 10.1103/PhysRevLett.118.240401} {\bibfield
  {journal} {\bibinfo  {journal} {Phys. Rev. Lett.}\ }\textbf {\bibinfo
  {volume} {118}},\ \bibinfo {pages} {240401} (\bibinfo {year}
  {2017})}\BibitemShut {NoStop}%
\bibitem [{\citenamefont {Virz\`{\i}}\ \emph {et~al.}(2022)\citenamefont
  {Virz\`{\i}}, \citenamefont {Avella}, \citenamefont {Piacentini},
  \citenamefont {Gramegna}, \citenamefont {Opatrn\'y}, \citenamefont {Kofman},
  \citenamefont {Kurizki}, \citenamefont {Gherardini}, \citenamefont {Caruso},
  \citenamefont {Degiovanni},\ and\ \citenamefont {Genovese}}]{Virzi2022}%
  \BibitemOpen
  {\bibfield  {author} {\bibinfo {author} {\bibfnamefont {S.}~\bibnamefont
  {Virz\`{\i}}}, \bibinfo {author} {\bibfnamefont {A.}~\bibnamefont {Avella}},
  \bibinfo {author} {\bibfnamefont {F.}~\bibnamefont {Piacentini}}, \bibinfo
  {author} {\bibfnamefont {M.}~\bibnamefont {Gramegna}}, \bibinfo {author}
  {\bibfnamefont {T.}~\bibnamefont {Opatrn\'y}}, \bibinfo {author}
  {\bibfnamefont {A.~G.}\ \bibnamefont {Kofman}}, \bibinfo {author}
  {\bibfnamefont {G.}~\bibnamefont {Kurizki}}, \bibinfo {author} {\bibfnamefont
  {S.}~\bibnamefont {Gherardini}}, \bibinfo {author} {\bibfnamefont
  {F.}~\bibnamefont {Caruso}}, \bibinfo {author} {\bibfnamefont {I.~P.}\
  \bibnamefont {Degiovanni}}, \ and\ \bibinfo {author} {\bibfnamefont
  {M.}~\bibnamefont {Genovese}},\ }\bibfield  {title} {\bibinfo {title} {Quantum Zeno and anti-Zeno probes of noise correlations in photon polarization},\ }\href {\doibase
  10.1103/PhysRevLett.129.030401} {\bibfield  {journal} {\bibinfo  {journal}
  {Phys. Rev. Lett.}\ }\textbf {\bibinfo {volume} {129}},\ \bibinfo {pages}
  {030401} (\bibinfo {year} {2022})}}\BibitemShut {NoStop}%
\bibitem [{\citenamefont {Kiilerich}\ and\ \citenamefont
  {M\o{}lmer}(2015)}]{Kiilerich2015}%
  \BibitemOpen
  {\bibfield  {author} {\bibinfo {author} {\bibfnamefont {A.~H.}\ \bibnamefont
  {Kiilerich}}\ and\ \bibinfo {author} {\bibfnamefont {K.}~\bibnamefont
  {M\o{}lmer}},\ }\bibfield  {title} {\bibinfo {title} {Quantum Zeno effect in parameter estimation},\ }\href {\doibase 10.1103/PhysRevA.92.032124} {\bibfield
  {journal} {\bibinfo  {journal} {Phys. Rev. A}\ }\textbf {\bibinfo {volume}
  {92}},\ \bibinfo {pages} {032124} (\bibinfo {year} {2015})}}\BibitemShut
  {NoStop}%
\bibitem [{\citenamefont {Do}\ \emph {et~al.}(2019)\citenamefont {Do},
  \citenamefont {Lovecchio}, \citenamefont {Mastroserio}, \citenamefont
  {Fabbri}, \citenamefont {Cataliotti}, \citenamefont {Gherardini},
  \citenamefont {M\"{u}ller}, \citenamefont {Pozza},\ and\ \citenamefont
  {Caruso}}]{Do2019}%
  \BibitemOpen
  {\bibfield  {author} {\bibinfo {author} {\bibfnamefont {H.-V.}\ \bibnamefont
  {Do}}, \bibinfo {author} {\bibfnamefont {C.}~\bibnamefont {Lovecchio}},
  \bibinfo {author} {\bibfnamefont {I.}~\bibnamefont {Mastroserio}}, \bibinfo
  {author} {\bibfnamefont {N.}~\bibnamefont {Fabbri}}, \bibinfo {author}
  {\bibfnamefont {F.~S.}\ \bibnamefont {Cataliotti}}, \bibinfo {author}
  {\bibfnamefont {S.}~\bibnamefont {Gherardini}}, \bibinfo {author}
  {\bibfnamefont {M.~M.}\ \bibnamefont {M\"{u}ller}}, \bibinfo {author}
  {\bibfnamefont {N.~D.}\ \bibnamefont {Pozza}}, \ and\ \bibinfo {author}
  {\bibfnamefont {F.}~\bibnamefont {Caruso}},\ }\bibfield  {title} {\bibinfo {title} {Experimental proof of quantum Zeno-assisted noise sensing},\ }\href {\doibase
  10.1088/1367-2630/ab5740} {\bibfield  {journal} {\bibinfo  {journal} {New J.
  Phys.}\ }\textbf {\bibinfo {volume} {21}},\ \bibinfo {pages} {113056}
  (\bibinfo {year} {2019})}}\BibitemShut {NoStop}%
\bibitem [{\citenamefont {Burgarth}\ \emph {et~al.}(2014)\citenamefont
  {Burgarth}, \citenamefont {Facchi}, \citenamefont {Giovannetti},
  \citenamefont {Nakazato}, \citenamefont {Pascazio},\ and\ \citenamefont
  {Yuasa}}]{Burgarth2014}%
  \BibitemOpen
  {\bibfield  {author} {\bibinfo {author} {\bibfnamefont {D.~K.}\ \bibnamefont
  {Burgarth}}, \bibinfo {author} {\bibfnamefont {P.}~\bibnamefont {Facchi}},
  \bibinfo {author} {\bibfnamefont {V.}~\bibnamefont {Giovannetti}}, \bibinfo
  {author} {\bibfnamefont {H.}~\bibnamefont {Nakazato}}, \bibinfo {author}
  {\bibfnamefont {S.}~\bibnamefont {Pascazio}}, \ and\ \bibinfo {author}
  {\bibfnamefont {K.}~\bibnamefont {Yuasa}},\ }\bibfield  {title} {\bibinfo {title} {Exponential rise of dynamical complexity in quantum computing through projections},\ }\href {\doibase
  10.1038/ncomms6173} {\bibfield  {journal} {\bibinfo  {journal} {Nat.
  Commun.}\ }\textbf {\bibinfo {volume} {5}},\ \bibinfo {pages} {5173}
  (\bibinfo {year} {2014})}}\BibitemShut {NoStop}%
\bibitem [{\citenamefont {Chin}\ \emph {et~al.}(2012)\citenamefont {Chin},
  \citenamefont {Huelga},\ and\ \citenamefont {Plenio}}]{Chin2012}%
  \BibitemOpen
  \bibfield  {author} {\bibinfo {author} {\bibfnamefont {A.~W.}\ \bibnamefont
  {Chin}}, \bibinfo {author} {\bibfnamefont {S.~F.}\ \bibnamefont {Huelga}}, \
  and\ \bibinfo {author} {\bibfnamefont {M.~B.}\ \bibnamefont {Plenio}},\
  }\bibfield  {title} {\bibinfo {title} {Quantum metrology in non-Markovian
  environments},\ }\href {\doibase 10.1103/PhysRevLett.109.233601} {\bibfield  {journal}
  {\bibinfo  {journal} {Phys. Rev. Lett.}\ }\textbf {\bibinfo {volume} {109}},\
  \bibinfo {pages} {233601} (\bibinfo {year} {2012})}\BibitemShut {NoStop}%
\bibitem [{\citenamefont {Matsuzaki}\ \emph {et~al.}(2011)\citenamefont
  {Matsuzaki}, \citenamefont {Benjamin},\ and\ \citenamefont
  {Fitzsimons}}]{Matsuzaki2011}%
  \BibitemOpen
  \bibfield  {author} {\bibinfo {author} {\bibfnamefont {Y.}~\bibnamefont
  {Matsuzaki}}, \bibinfo {author} {\bibfnamefont {S.~C.}\ \bibnamefont
  {Benjamin}}, \ and\ \bibinfo {author} {\bibfnamefont {J.}~\bibnamefont
  {Fitzsimons}},\ }\bibfield  {title} {\bibinfo {title} {Magnetic field sensing
  beyond the standard quantum limit under the effect of decoherence},\
  }\href {\doibase 10.1103/PhysRevA.84.012103} {\bibfield
  {journal} {\bibinfo  {journal} {Phys. Rev. A}\ }\textbf {\bibinfo {volume}
  {84}},\ \bibinfo {pages} {012103} (\bibinfo {year} {2011})}\BibitemShut
  {NoStop}%
\bibitem [{\citenamefont {Wang}\ \emph {et~al.}(2018)\citenamefont {Wang},
  \citenamefont {Tao}, \citenamefont {Ai}, \citenamefont {Xin}, \citenamefont
  {Lambert}, \citenamefont {Ruan}, \citenamefont {Cheng}, \citenamefont {Nori},
  \citenamefont {Deng},\ and\ \citenamefont {Long}}]{Wang2018b}%
  \BibitemOpen
  \bibfield  {author} {\bibinfo {author} {\bibfnamefont {B.-X.}\ \bibnamefont
  {Wang}}, \bibinfo {author} {\bibfnamefont {M.-J.}\ \bibnamefont {Tao}},
  \bibinfo {author} {\bibfnamefont {Q.}~\bibnamefont {Ai}}, \bibinfo {author}
  {\bibfnamefont {T.}~\bibnamefont {Xin}}, \bibinfo {author} {\bibfnamefont
  {N.}~\bibnamefont {Lambert}}, \bibinfo {author} {\bibfnamefont
  {D.}~\bibnamefont {Ruan}}, \bibinfo {author} {\bibfnamefont {Y.-C.}\
  \bibnamefont {Cheng}}, \bibinfo {author} {\bibfnamefont {F.}~\bibnamefont
  {Nori}}, \bibinfo {author} {\bibfnamefont {F.-G.}\ \bibnamefont {Deng}}, \
  and\ \bibinfo {author} {\bibfnamefont {G.-L.}\ \bibnamefont {Long}},\ }\bibfield  {title} {\bibinfo {title} {Efficient quantum simulation of photosynthetic light harvesting},\ }\href
  {\doibase 10.1038/s41534-018-0102-2} {\bibfield  {journal} {\bibinfo
  {journal} {npj Quantum Inf.}\ }\textbf {\bibinfo {volume} {4}},\ \bibinfo
  {pages} {52} (\bibinfo {year} {2018})}\BibitemShut {NoStop}%
\bibitem [{\citenamefont {Zhang}\ \emph {et~al.}(2021)\citenamefont {Zhang},
  \citenamefont {Tao}, \citenamefont {He}, \citenamefont {Chen}, \citenamefont
  {Kong}, \citenamefont {Deng}, \citenamefont {Lambert},\ and\ \citenamefont
  {Ai}}]{Zhang2021}%
  \BibitemOpen
  \bibfield  {author} {\bibinfo {author} {\bibfnamefont {N.-N.}\ \bibnamefont
  {Zhang}}, \bibinfo {author} {\bibfnamefont {M.-J.}\ \bibnamefont {Tao}},
  \bibinfo {author} {\bibfnamefont {W.-T.}\ \bibnamefont {He}}, \bibinfo
  {author} {\bibfnamefont {X.-Y.}\ \bibnamefont {Chen}}, \bibinfo {author}
  {\bibfnamefont {X.-Y.}\ \bibnamefont {Kong}}, \bibinfo {author}
  {\bibfnamefont {F.-G.}\ \bibnamefont {Deng}}, \bibinfo {author}
  {\bibfnamefont {N.}~\bibnamefont {Lambert}}, \ and\ \bibinfo {author}
  {\bibfnamefont {Q.}~\bibnamefont {Ai}},\ }\bibfield  {title} {\bibinfo
  {title} {Efficient quantum simulation of open quantum dynamics at various
  Hamiltonians and spectral densities},\ }\href {\doibase
  10.1007/s11467-021-1064-y} {\bibfield  {journal} {\bibinfo  {journal} {Front.
  Phys.}\ }\textbf {\bibinfo {volume} {16}},\ \bibinfo {pages} {51501}
  (\bibinfo {year} {2021})}\BibitemShut {NoStop}%
\bibitem [{\citenamefont {Chen}\ \emph {et~al.}(2022)\citenamefont {Chen},
  \citenamefont {Zhang}, \citenamefont {He}, \citenamefont {Kong},
  \citenamefont {Tao}, \citenamefont {Deng}, \citenamefont {Ai},\ and\
  \citenamefont {Long}}]{Chen2022}%
  \BibitemOpen
  \bibfield  {author} {\bibinfo {author} {\bibfnamefont {X.-Y.}\ \bibnamefont
  {Chen}}, \bibinfo {author} {\bibfnamefont {N.-N.}\ \bibnamefont {Zhang}},
  \bibinfo {author} {\bibfnamefont {W.-T.}\ \bibnamefont {He}}, \bibinfo
  {author} {\bibfnamefont {X.-Y.}\ \bibnamefont {Kong}}, \bibinfo {author}
  {\bibfnamefont {M.-J.}\ \bibnamefont {Tao}}, \bibinfo {author} {\bibfnamefont
  {F.-G.}\ \bibnamefont {Deng}}, \bibinfo {author} {\bibfnamefont
  {Q.}~\bibnamefont {Ai}}, \ and\ \bibinfo {author} {\bibfnamefont {G.-L.}\
  \bibnamefont {Long}},\ }\bibfield  {title} {\bibinfo
  {title} {Global correlation and local information flows in controllable
  non-Markovian open quantum dynamics},\ }\href {\doibase 10.1038/s41534-022-00537-z}
  {\bibfield  {journal} {\bibinfo  {journal} {npj Quantum Inf.}\ }\textbf
  {\bibinfo {volume} {8}},\ \bibinfo {pages} {22} (\bibinfo {year}
  {2022})}\BibitemShut {NoStop}%
\bibitem [{\citenamefont {Soare}\ \emph
  {et~al.}(2014{\natexlab{a}})\citenamefont {Soare}, \citenamefont {Ball},
  \citenamefont {Hayes}, \citenamefont {Sastrawan}, \citenamefont {Jarratt},
  \citenamefont {McLoughlin}, \citenamefont {Zhen}, \citenamefont {Green},\
  and\ \citenamefont {Biercuk}}]{Soare2014a}%
  \BibitemOpen
  \bibfield  {author} {\bibinfo {author} {\bibfnamefont {A.}~\bibnamefont
  {Soare}}, \bibinfo {author} {\bibfnamefont {H.}~\bibnamefont {Ball}},
  \bibinfo {author} {\bibfnamefont {D.}~\bibnamefont {Hayes}}, \bibinfo
  {author} {\bibfnamefont {J.}~\bibnamefont {Sastrawan}}, \bibinfo {author}
  {\bibfnamefont {M.~C.}\ \bibnamefont {Jarratt}}, \bibinfo {author}
  {\bibfnamefont {J.~J.}\ \bibnamefont {McLoughlin}}, \bibinfo {author}
  {\bibfnamefont {X.}~\bibnamefont {Zhen}}, \bibinfo {author} {\bibfnamefont
  {T.~J.}\ \bibnamefont {Green}}, \ and\ \bibinfo {author} {\bibfnamefont
  {M.~J.}\ \bibnamefont {Biercuk}},\ }\bibfield  {title} {\bibinfo {title}
  {Experimental noise filtering by quantum control},\ }\href {\doibase 10.1038/nphys3115}
  {\bibfield  {journal} {\bibinfo  {journal} {Nat. Phys.}\ }\textbf {\bibinfo
  {volume} {10}},\ \bibinfo {pages} {825} (\bibinfo {year}
  {2014}{\natexlab{a}})}\BibitemShut {NoStop}%
\bibitem [{\citenamefont {Soare}\ \emph
  {et~al.}(2014{\natexlab{b}})\citenamefont {Soare}, \citenamefont {Ball},
  \citenamefont {Hayes}, \citenamefont {Zhen}, \citenamefont {Jarratt},
  \citenamefont {Sastrawan}, \citenamefont {Uys},\ and\ \citenamefont
  {Biercuk}}]{Soare2014}%
  \BibitemOpen
  \bibfield  {author} {\bibinfo {author} {\bibfnamefont {A.}~\bibnamefont
  {Soare}}, \bibinfo {author} {\bibfnamefont {H.}~\bibnamefont {Ball}},
  \bibinfo {author} {\bibfnamefont {D.}~\bibnamefont {Hayes}}, \bibinfo
  {author} {\bibfnamefont {X.}~\bibnamefont {Zhen}}, \bibinfo {author}
  {\bibfnamefont {M.~C.}\ \bibnamefont {Jarratt}}, \bibinfo {author}
  {\bibfnamefont {J.}~\bibnamefont {Sastrawan}}, \bibinfo {author}
  {\bibfnamefont {H.}~\bibnamefont {Uys}}, \ and\ \bibinfo {author}
  {\bibfnamefont {M.~J.}\ \bibnamefont {Biercuk}},\ }\bibfield  {title}
  {\bibinfo {title} {Experimental bath engineering for quantitative studies of
  quantum control},\ }\href {\doibase
  10.1103/PhysRevA.89.042329} {\bibfield  {journal} {\bibinfo  {journal} {Phys.
  Rev. A}\ }\textbf {\bibinfo {volume} {89}},\ \bibinfo {pages} {042329}
  (\bibinfo {year} {2014}{\natexlab{b}})}\BibitemShut {NoStop}%
\bibitem [{\citenamefont {Khaneja}\ \emph {et~al.}(2005)\citenamefont
  {Khaneja}, \citenamefont {Reiss}, \citenamefont {Kehlet}, \citenamefont
  {{Schulte-Herbr{\"u}ggen}},\ and\ \citenamefont {Glaser}}]{Khaneja2005b}%
  \BibitemOpen
  \bibfield  {author} {\bibinfo {author} {\bibfnamefont {N.}~\bibnamefont
  {Khaneja}}, \bibinfo {author} {\bibfnamefont {T.}~\bibnamefont {Reiss}},
  \bibinfo {author} {\bibfnamefont {C.}~\bibnamefont {Kehlet}}, \bibinfo
  {author} {\bibfnamefont {T.}~\bibnamefont {Schulte-Herbr{\"u}ggen}}, \ }\bibfield  {title} {\bibinfo {title} {Optimal control of coupled spin dynamics: Design of NMR pulse sequences by gradient ascent algorithms},\ }\href
  {\doibase 10.1016/j.jmr.2004.11.004} {\bibfield  {journal} {\bibinfo
  {journal} {J. Magn. Reson. }\ }\textbf {\bibinfo {volume}
  {172}},\ \bibinfo {pages} {296} (\bibinfo {year} {2005})}\BibitemShut
  {NoStop}%
\bibitem [{\citenamefont {Li}\ \emph {et~al.}(2017)\citenamefont {Li},
  \citenamefont {Yang}, \citenamefont {Peng},\ and\ \citenamefont
  {Sun}}]{Li2017}%
  \BibitemOpen
  \bibfield  {author} {\bibinfo {author} {\bibfnamefont {J.}~\bibnamefont
  {Li}}, \bibinfo {author} {\bibfnamefont {X.}~\bibnamefont {Yang}}, \bibinfo
  {author} {\bibfnamefont {X.}~\bibnamefont {Peng}}, \ and\ \bibinfo {author}
  {\bibfnamefont {C.-P.}\ \bibnamefont {Sun}},\ }\bibfield
  {title} {\bibinfo {title} {Hybrid quantum-classical approach to quantum
  optimal control},\ }\href {\doibase
  10.1103/PhysRevLett.118.150503} {\bibfield  {journal} {\bibinfo  {journal}
  {Phys. Rev. Lett.}\ }\textbf {\bibinfo {volume} {118}},\ \bibinfo {pages}
  {150503} (\bibinfo {year} {2017})}\BibitemShut {NoStop}%
\bibitem [{\citenamefont {Buluta}\ and\ \citenamefont
  {Nori}(2009)}]{Buluta2009}%
  \BibitemOpen
  \bibfield  {author} {\bibinfo {author} {\bibfnamefont {I.}~\bibnamefont
  {Buluta}}\ and\ \bibinfo {author} {\bibfnamefont {F.}~\bibnamefont {Nori}},\
  }\bibfield
  {title} {\bibinfo {title} {Quantum simulators},\ }\href {\doibase 10.1126/science.1177838} {\bibfield  {journal} {\bibinfo
  {journal} {Science}\ }\textbf {\bibinfo {volume} {326}},\ \bibinfo {pages}
  {108} (\bibinfo {year} {2009})}\BibitemShut {NoStop}%
\bibitem [{\citenamefont {Georgescu}\ \emph {et~al.}(2014)\citenamefont
  {Georgescu}, \citenamefont {Ashhab},\ and\ \citenamefont
  {Nori}}]{Georgescu2014}%
  \BibitemOpen
  \bibfield  {author} {\bibinfo {author} {\bibfnamefont {I.~M.}\ \bibnamefont
  {Georgescu}}, \bibinfo {author} {\bibfnamefont {S.}~\bibnamefont {Ashhab}}, \
  and\ \bibinfo {author} {\bibfnamefont {F.}~\bibnamefont {Nori}},\ }\bibfield
  {title} {\bibinfo {title} {Quantum simulation},\ }\href
  {\doibase 10.1103/RevModPhys.86.153} {\bibfield  {journal} {\bibinfo
  {journal} {Rev. Mod. Phys.}\ }\textbf {\bibinfo {volume} {86}},\ \bibinfo
  {pages} {153} (\bibinfo {year} {2014})}\BibitemShut {NoStop}%
\bibitem [{\citenamefont {Luo}\ \emph {et~al.}(2012)\citenamefont {Luo},
  \citenamefont {Fu},\ and\ \citenamefont {Song}}]{Luo2012}%
  \BibitemOpen
  \bibfield  {author} {\bibinfo {author} {\bibfnamefont {S.}~\bibnamefont
  {Luo}}, \bibinfo {author} {\bibfnamefont {S.}~\bibnamefont {Fu}}, \ and\
  \bibinfo {author} {\bibfnamefont {H.}~\bibnamefont {Song}},\ }\bibfield
  {title} {\bibinfo {title} {Quantifying non-Markovianity via correlations},\
  }\href {\doibase
  10.1103/PhysRevA.86.044101} {\bibfield  {journal} {\bibinfo  {journal} {Phys.
  Rev. A}\ }\textbf {\bibinfo {volume} {86}},\ \bibinfo {pages} {044101}
  (\bibinfo {year} {2012})}\BibitemShut {NoStop}%
\bibitem [{\citenamefont {Lu}\ \emph {et~al.}(2010)\citenamefont {Lu},
  \citenamefont {Wang},\ and\ \citenamefont {Sun}}]{Lu2010}%
  \BibitemOpen
  \bibfield  {author} {\bibinfo {author} {\bibfnamefont {X.-M.}\ \bibnamefont
  {Lu}}, \bibinfo {author} {\bibfnamefont {X.}~\bibnamefont {Wang}}, \ and\
  \bibinfo {author} {\bibfnamefont {C.~P.}\ \bibnamefont {Sun}},\ }\bibfield
  {title} {\bibinfo {title} {Quantum Fisher information flow and non-Markovian
  processes of open systems},\ }\href
  {\doibase 10.1103/PhysRevA.82.042103} {\bibfield  {journal} {\bibinfo
  {journal} {Phys. Rev. A}\ }\textbf {\bibinfo {volume} {82}},\ \bibinfo
  {pages} {042103} (\bibinfo {year} {2010})}\BibitemShut {NoStop}%
\bibitem [{\citenamefont {Altherr}\ and\ \citenamefont
  {Yang}(2021)}]{Altherr2021}%
  \BibitemOpen
  {\bibfield  {author} {\bibinfo {author} {\bibfnamefont {A.}~\bibnamefont
  {Altherr}}\ and\ \bibinfo {author} {\bibfnamefont {Y.}~\bibnamefont {Yang}},\
  }\bibfield  {title} {\bibinfo {title} {Quantum metrology for non-Markovian processes},\ }\href {\doibase 10.1103/PhysRevLett.127.060501} {\bibfield  {journal}
  {\bibinfo  {journal} {Phys. Rev. Lett.}\ }\textbf {\bibinfo {volume} {127}},\
  \bibinfo {pages} {060501} (\bibinfo {year} {2021})}}\BibitemShut {NoStop}%
\bibitem [{\citenamefont {Goodman}(2015)}]{Goodman2015}%
  \BibitemOpen
  \bibfield  {author} {\bibinfo {author} {\bibfnamefont {J.~W.}\ \bibnamefont
  {Goodman}},\ }\href@noop {} {\emph {\bibinfo {title} {Statistical
  {{Optics}}}}}\ (\bibinfo  {publisher} {{John Wiley \& Sons}},\ \bibinfo
  {year} {2015})\BibitemShut {NoStop}%
\bibitem [{\citenamefont {Tao}\ \emph {et~al.}(2020)\citenamefont {Tao},
  \citenamefont {Zhang}, \citenamefont {Wen}, \citenamefont {Deng},
  \citenamefont {Ai},\ and\ \citenamefont {Long}}]{Tao2020}%
  \BibitemOpen
  \bibfield  {author} {\bibinfo {author} {\bibfnamefont {M.-J.}\ \bibnamefont
  {Tao}}, \bibinfo {author} {\bibfnamefont {N.-N.}\ \bibnamefont {Zhang}},
  \bibinfo {author} {\bibfnamefont {P.-Y.}\ \bibnamefont {Wen}}, \bibinfo
  {author} {\bibfnamefont {F.-G.}\ \bibnamefont {Deng}}, \bibinfo {author}
  {\bibfnamefont {Q.}~\bibnamefont {Ai}}, \ and\ \bibinfo {author}
  {\bibfnamefont {G.-L.}\ \bibnamefont {Long}},\ }\bibfield  {title} {\bibinfo
  {title} {Coherent and incoherent theories for photosynthetic energy
  transfer},\ }\href {\doibase
  10.1016/j.scib.2019.12.009} {\bibfield  {journal} {\bibinfo  {journal} {Sci.
  Bull.}\ }\textbf {\bibinfo {volume} {65}},\ \bibinfo {pages} {318} (\bibinfo
  {year} {2020})}\BibitemShut {NoStop}%
\bibitem [{sup()}]{sup}%
  \BibitemOpen
  \href@noop {} {\bibinfo  {journal} {See the supplemental information for the
  complete description of the theory, experimental details and numerical
  simulations}\ }\BibitemShut {NoStop}%
\bibitem [{\citenamefont {Knill}\ \emph {et~al.}(2000)\citenamefont {Knill},
  \citenamefont {Laflamme}, \citenamefont {Martinez},\ and\ \citenamefont
  {Tseng}}]{Knill2000a}%
  \BibitemOpen
\bibfield  {journal} {  }\bibfield  {author} {\bibinfo {author} {\bibfnamefont
  {E.}~\bibnamefont {Knill}}, \bibinfo {author} {\bibfnamefont
  {R.}~\bibnamefont {Laflamme}}, \bibinfo {author} {\bibfnamefont
  {R.}~\bibnamefont {Martinez}}, \ and\ \bibinfo {author} {\bibfnamefont
  {C.~H.}~\bibnamefont {Tseng}},\ }\bibfield  {title} {\bibinfo {title} {An
  algorithmic benchmark for quantum information processing},\ }\href {\doibase 10.1038/35006012} {\bibfield
  {journal} {\bibinfo  {journal} {Nature}\ }\textbf {\bibinfo {volume} {404}},\
  \bibinfo {pages} {368} (\bibinfo {year} {2000})}\BibitemShut {NoStop}%
\bibitem [{\citenamefont {Laflamme}\ \emph {et~al.}(2002)\citenamefont
  {Laflamme}, \citenamefont {Cory}, \citenamefont {Negrevergne},\ and\
  \citenamefont {Viola}}]{Laflamme2002a}%
  \BibitemOpen
  \bibfield  {author} {\bibinfo {author} {\bibfnamefont {R.}~\bibnamefont
  {Laflamme}}, \bibinfo {author} {\bibfnamefont {D.}~\bibnamefont {Cory}},
  \bibinfo {author} {\bibfnamefont {C.}~\bibnamefont {Negrevergne}}, \ and\
  \bibinfo {author} {\bibfnamefont {L.}~\bibnamefont {Viola}},\ }\bibfield
  {title} {\bibinfo {title} {{{NMR}} quantum information processing and
  entanglement},\ }\href
  {\doibase 10.5555/2011422.2011427} {\bibfield  {journal} {\bibinfo  {journal}
  {Quantum Inf. Comput.}\ }\textbf {\bibinfo {volume} {2}},\ \bibinfo
  {pages} {166} (\bibinfo {year} {2002})}\BibitemShut {NoStop}%
\bibitem [{\citenamefont {Poto{\v c}nik}\ \emph {et~al.}(2018)\citenamefont
  {Poto{\v c}nik}, \citenamefont {Bargerbos}, \citenamefont {Schr{\"o}der},
  \citenamefont {Khan}, \citenamefont {Collodo}, \citenamefont {Gasparinetti},
  \citenamefont {Salath{\'e}}, \citenamefont {Creatore}, \citenamefont
  {Eichler}, \citenamefont {T{\"u}reci}, \citenamefont {Chin},\ and\
  \citenamefont {Wallraff}}]{Potocnik2018}%
  \BibitemOpen
  \bibfield  {author} {\bibinfo {author} {\bibfnamefont {A.}~\bibnamefont
  {Poto{\v c}nik}}, \bibinfo {author} {\bibfnamefont {A.}~\bibnamefont
  {Bargerbos}}, \bibinfo {author} {\bibfnamefont {F.~A. Y.~N.}\ \bibnamefont
  {Schr{\"o}der}}, \bibinfo {author} {\bibfnamefont {S.~A.}\ \bibnamefont
  {Khan}}, \bibinfo {author} {\bibfnamefont {M.~C.}\ \bibnamefont {Collodo}},
  \bibinfo {author} {\bibfnamefont {S.}~\bibnamefont {Gasparinetti}}, \bibinfo
  {author} {\bibfnamefont {Y.}~\bibnamefont {Salath{\'e}}}, \bibinfo {author}
  {\bibfnamefont {C.}~\bibnamefont {Creatore}}, \bibinfo {author}
  {\bibfnamefont {C.}~\bibnamefont {Eichler}}, \bibinfo {author} {\bibfnamefont
  {H.~E.}\ \bibnamefont {T{\"u}reci}}, \bibinfo {author} {\bibfnamefont
  {A.~W.}\ \bibnamefont {Chin}}, \ and\ \bibinfo {author} {\bibfnamefont
  {A.}~\bibnamefont {Wallraff}},\ }\bibfield  {title} {\bibinfo {title} {Studying light-harvesting models with superconducting circuits},\ }\href {\doibase 10.1038/s41467-018-03312-x} {\bibfield  {journal} {\bibinfo  {journal} {Nat. Commun}\ }\textbf {\bibinfo
  {volume} {9}},\ \bibinfo {pages} {904} (\bibinfo {year} {2018})}\BibitemShut
  {NoStop}%
\end{thebibliography}

\end{document}